\documentclass[10pt,journal,compsoc]{IEEEtran}
\usepackage{graphicx}
\usepackage{hyperref}
\usepackage{amssymb}
\usepackage{amsmath}
\usepackage{amsthm}
\usepackage{makecell}
\usepackage{multicol}
\usepackage{multirow}
\usepackage{booktabs}
\usepackage{pifont}
\usepackage{bm}
\usepackage{color}
\usepackage{enumitem}
\usepackage[table]{xcolor}
\ifCLASSOPTIONcompsoc
  \usepackage[nocompress]{cite}
\else
  \usepackage{cite}
\fi
\ifCLASSINFOpdf
\else
\fi
\begin{document}
\title{An Efficient High-Degree, High-Order Equivariant Graph Neural Network for Direct Crystal Structure Optimization}

\author{Ziduo Yang,
        Wei Zhuo,
        Huiqiang Xie,
        Xiaoqing Liu,
        and Lei Shen% <-this % stops a space
\IEEEcompsocitemizethanks{\IEEEcompsocthanksitem This research was supported by the National Natural Science Foundation of China (Grant No. 62506143), the Natural Science Foundation of Guangdong Province (Grant No. 2025A1515011487), Ministry of Education, Singapore, Tier 1 (Grant No. A-8001194-00-00), and Tier 2  (Grant No. A-8001872-00-00). (Corresponding Author: Huiqiang Xie and Xiaoqing Liu.)

\IEEEcompsocthanksitem Ziduo Yang and Huiqiang Xie are with the College of Information Science and Technology, Jinan University, Guangzhou, 510632, China. Wei Zhuo is with the College of Computing and Data Science, Nanyang Technological University, 50 Nanyang Avenue, 639798, Singapore. Xiaoqing Liu and Lei Shen are with the Department of Mechanical Engineering, National University of Singapore, 9 Engineering Drive 1, 117575, Singapore. Lei Shen is also with the  National University of Singapore (Chongqing) Research Institute, Chongqing, 401123, China.
}% <-this % stops an unwanted space
}


\markboth{Journal of \LaTeX\ Class Files,~Vol.~14, No.~8, August~2015}%
{Shell \MakeLowercase{\textit{et al.}}: Bare Demo of IEEEtran.cls for Computer Society Journals}



\IEEEtitleabstractindextext{%
\begin{abstract}
Crystal structure optimization is fundamental to materials modeling but remains computationally expensive when performed with density-functional theory (DFT). Machine-learning (ML) approaches offer substantial acceleration, yet existing methods face three key limitations: (i) most models operate solely on atoms and treat lattice vectors implicitly, despite their central role in structural optimization; (ii) they lack efficient mechanisms to capture high-degree angular information and higher-order geometric correlations simultaneously, which are essential for distinguishing subtle structural differences; and (iii) many pipelines are multi-stage or iterative rather than truly end-to-end, making them prone to error accumulation and limiting scalability. Here we present E$^{3}$Relax-H$^{2}$, an end-to-end high-degree, high-order equivariant graph neural network that maps an initial crystal directly to its relaxed structure. The key idea is to promote both atoms and lattice vectors to graph nodes, enabling a unified and symmetry-consistent representation of structural degrees of freedom. Building on this formulation, E$^{3}$Relax-H$^{2}$ introduces two message-passing mechanisms: (i) a high-degree, high-order message-passing module that efficiently captures high-degree angular representations and high-order many-body correlations; and (ii) a lattice-atom message-passing module that explicitly models the bidirectional coupling between lattice deformation and atomic displacement. In addition, we propose a differentiable periodicity-aware Cartesian displacement loss tailored for one-shot structure prediction under periodic boundary conditions. We evaluate E$^{3}$Relax-H$^{2}$ on six benchmark datasets spanning diverse materials systems and covering both mild and severe structural distortions from the relaxed configuration. The results demonstrate that E$^{3}$Relax-H$^{2}$ achieves state-of-the-art or highly competitive structural accuracy while maintaining high computational efficiency. Additional DFT validation further confirms that the predicted geometries are energetically favorable. The source code is available at \texttt{https://github.com/guaguabujianle/E3RelaxH2}.
\end{abstract}

\begin{IEEEkeywords}
Graph Neural Networks, Geometric Deep Learning, Crystal Structure Optimization
\end{IEEEkeywords}}

\maketitle

\IEEEdisplaynontitleabstractindextext

% \fi
\IEEEpeerreviewmaketitle

\section{Introduction}

\begin{figure*}[!tb]
  \centering
  \includegraphics[width=2\columnwidth]{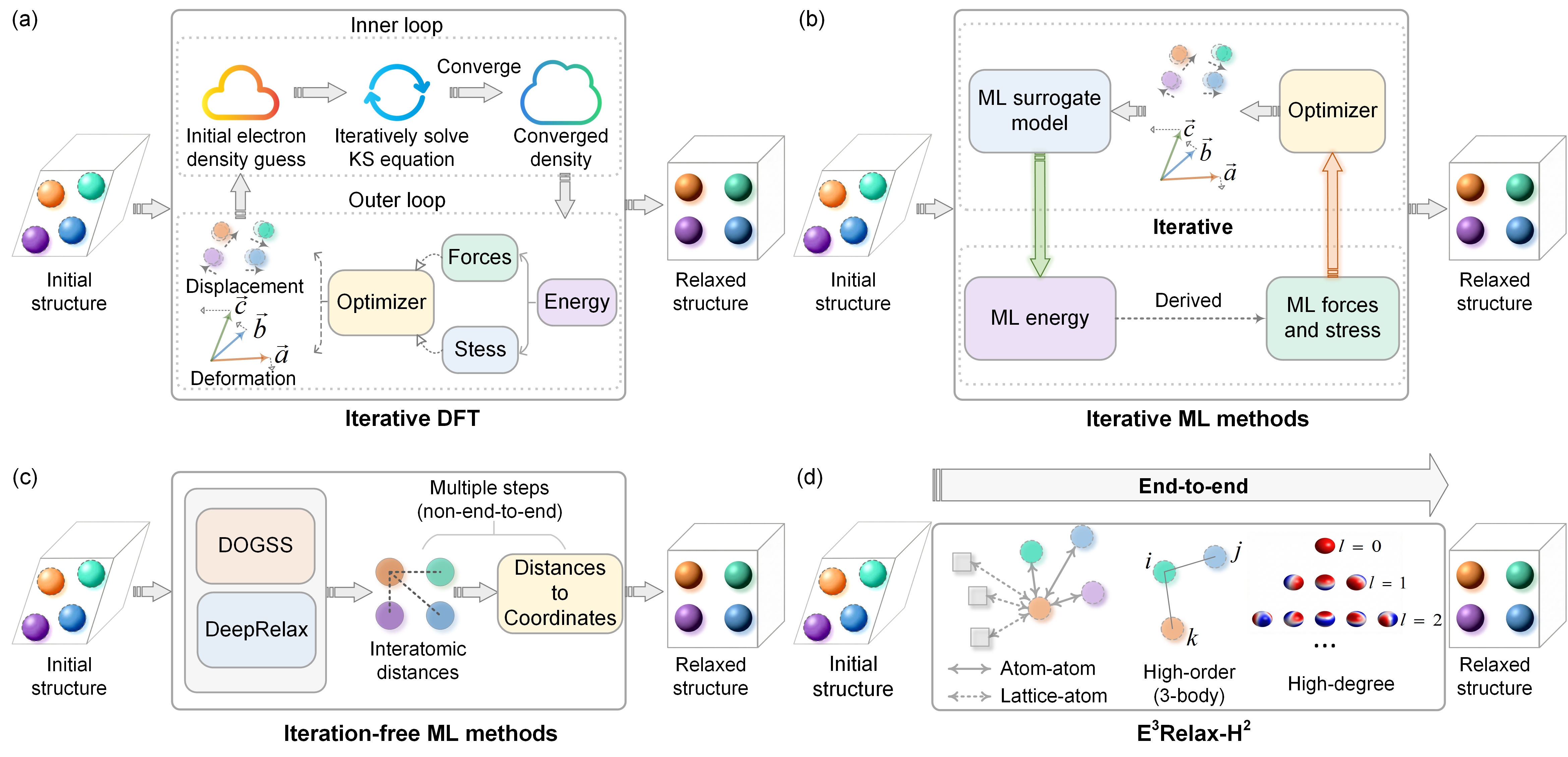}
  \caption{Comparison of structural optimization using DFT and machine learning.
  (a) Conventional DFT-based structural relaxation, which involves two nested loops: an inner self-consistent electronic optimization and an outer geometry update loop.
  (b) Iterative ML-based relaxation methods, where a learned force field replaces the DFT inner loop, but an explicit geometry-update loop is still required.
  (c) Existing iteration-free ML approaches that bypass iterative relaxation by first predicting relaxed interatomic distances and subsequently reconstructing atomic coordinates.
  (d) Our iteration-free, end-to-end structure optimization framework, which directly predicts atomic and lattice displacements under SE(3)-equivariance, enabling a single forward pass from the initial structure to the relaxed configuration without intermediate optimization steps.}
  \label{fgr:struct_relax}
\end{figure*}

Structural optimization, the process of driving a crystal structure to its minimum-energy configuration (commonly called the relaxed structure), is a fundamental step in computational materials science, as most material properties are evaluated based on the relaxed geometry. Traditionally, this task is carried out using \textit{ab initio} methods such as density functional theory (DFT) \cite{gibson2022data, lyngby2022data}. However, DFT-based relaxation is computationally expensive due to its two nested iterative loops (Figure \ref{fgr:struct_relax}(a)): the inner loop repeatedly solves the Kohn-Sham equations to self-consistency to obtain energies and forces, while the outer loop updates atomic coordinates until the forces fall below a convergence threshold. Together, these loops require substantial computation to reach the minimum-energy structure.

To mitigate the high computational cost of DFT-based structure optimization, machine learning (ML) has emerged as a promising alternative for efficiently predicting relaxed crystal structures. Existing ML approaches can be broadly categorized into iterative methods \cite{chen2022universal, deng2023chgnet, batatia2022mace, gasteiger_dimenet_2020, mosquera2024machine, jiang2024machine, yang2025efficient, liu2026traj2relax, lin2026learning} and iteration-free methods \cite{kim2023structure, yoon2020differentiable, yang2024scalable}.

Iterative ML methods replace the expensive DFT inner loop with a learned surrogate while retaining an outer geometry-update loop. A typical pipeline employs machine learning interatomic potentials (MLIPs), which predict energies, forces, and stresses at each step to guide structural updates (Figure~\ref{fgr:struct_relax}(b)). Although significantly faster than DFT, these approaches require dense supervision from DFT trajectories and remain inherently sequential due to the iterative optimization process.

In contrast, iteration-free methods aim to directly predict relaxed structures from unrelaxed inputs, eliminating iterative optimization. Representative approaches, such as DeepRelax \cite{yang2024scalable} and DOGSS \cite{yoon2020differentiable}, adopt a two-stage pipeline that predicts interatomic distances and then reconstructs coordinates via distance geometry (Figure~\ref{fgr:struct_relax}(c)). Despite their iteration-free advantages, these methods face three key limitations. First, they model only atomic interactions and neglect explicit lattice deformation. In practice, structural relaxation involves coupled atomic displacements and lattice parameter changes, and variations in cell geometry can substantially affect material properties. Second, they lack efficient mechanisms to capture high-fidelity angular information and high-order geometric correlations, both of which are critical for distinguishing subtle structural configurations. Third, their multi-stage pipelines are not fully end-to-end, leading to potential error accumulation.

To overcome these challenges, we propose E$^{3}$Relax-H$^2$, a high-degree equivariant graph neural network that predicts the relaxed crystal structure in a single forward pass (Figure \ref{fgr:struct_relax}(d)). E$^{3}$Relax-H$^2$ introduces four key advantages:

\begin{itemize}
    \item \textbf{Unified atomic and lattice modeling.} E$^{3}$Relax-H$^2$ promotes both atoms and lattice vectors to graph nodes, each equipped with an invariant scalar and an equivariant 3D vector feature. This dual-node design enables the network to jointly model atomic displacements and lattice deformations simultaneously, overcoming the limitation of earlier models that represent only atoms.
    \item \textbf{Efficient high-degree, high-order message passing (H$^2$-MP) module.} We introduce a novel message-passing mechanism that captures high-degree angular representations and many-body geometric correlations. This design preserves the expressive power needed to discriminate subtle structural differences while achieving significantly lower computational cost than traditional tensor-product-based equivariant formulations.
    \item \textbf{Lattice-atom message passing (LA-MP) module.} We design a module that explicitly captures the bidirectional coupling between atomic displacements and lattice deformations under the dual-node representation. This enables coherent propagation of structural information across atoms and lattice vectors, an effect overlooked in previous iteration-free methods.
    \item \textbf{Periodicity-aware direct structure supervision.} E$^{3}$Relax-H$^2$ directly predicts the relaxed crystal structure from the unrelaxed input and is trained with a differentiable periodicity-aware Cartesian displacement loss tailored for one-shot prediction under periodic boundary conditions. This design provides physically consistent supervision for direct structure optimization and reduces the error accumulation associated with multi-stage approaches.
\end{itemize}

This work is a substantial extension of our earlier conference paper, E$^{3}$Relax, published at AAAI-26 \cite{yang2026equivariant}, in terms of model design, supervision, and empirical evaluation. Compared with E$^{3}$Relax, the present work introduces four key advances:
\begin{itemize}
    \item \textbf{More expressive geometric modeling beyond E$^{3}$Relax (Section \ref{sec:h2mp}).} Compared with E$^{3}$Relax, we introduce an H$^2$-MP module that substantially strengthens geometric expressivity by simultaneously capturing richer angular information and implicit many-body correlations, while maintaining low computational overhead.

    \item \textbf{More physically consistent supervision (Section \ref{sec:pacd}).} Compared with the supervision strategy used in E$^{3}$Relax, we introduce a PACD loss tailored for one-shot crystal structure optimization under periodic boundary conditions. This novel loss provides more physically consistent supervision and improves predictive accuracy.

    \item \textbf{Translation-invariant lattice-atom coupling (Section \ref{sec:lamp}).} We redesign the lattice-atom interaction module to restore translation invariance. Compared with the earlier E$^{3}$Relax design, the novel formulation uses translation-consistent geometric descriptors for lattice--atom coupling, yielding a more physically grounded architecture and improving robustness across datasets with different structural characteristics.

    \item \textbf{Broader and deeper empirical validation.} We substantially broaden the empirical scope of the study by adding new experiments on the JARVIS benchmark \cite{choudhary2020joint}, the OC20 benchmark \cite{chanussot2021open}, and a newly constructed 2D materials dataset. We further provide a more comprehensive analysis of the proposed framework, offering deeper insight into the model's behavior and advantages.
\end{itemize}

\begin{figure*}[t]
  \centering
  \includegraphics[width=1.8\columnwidth]{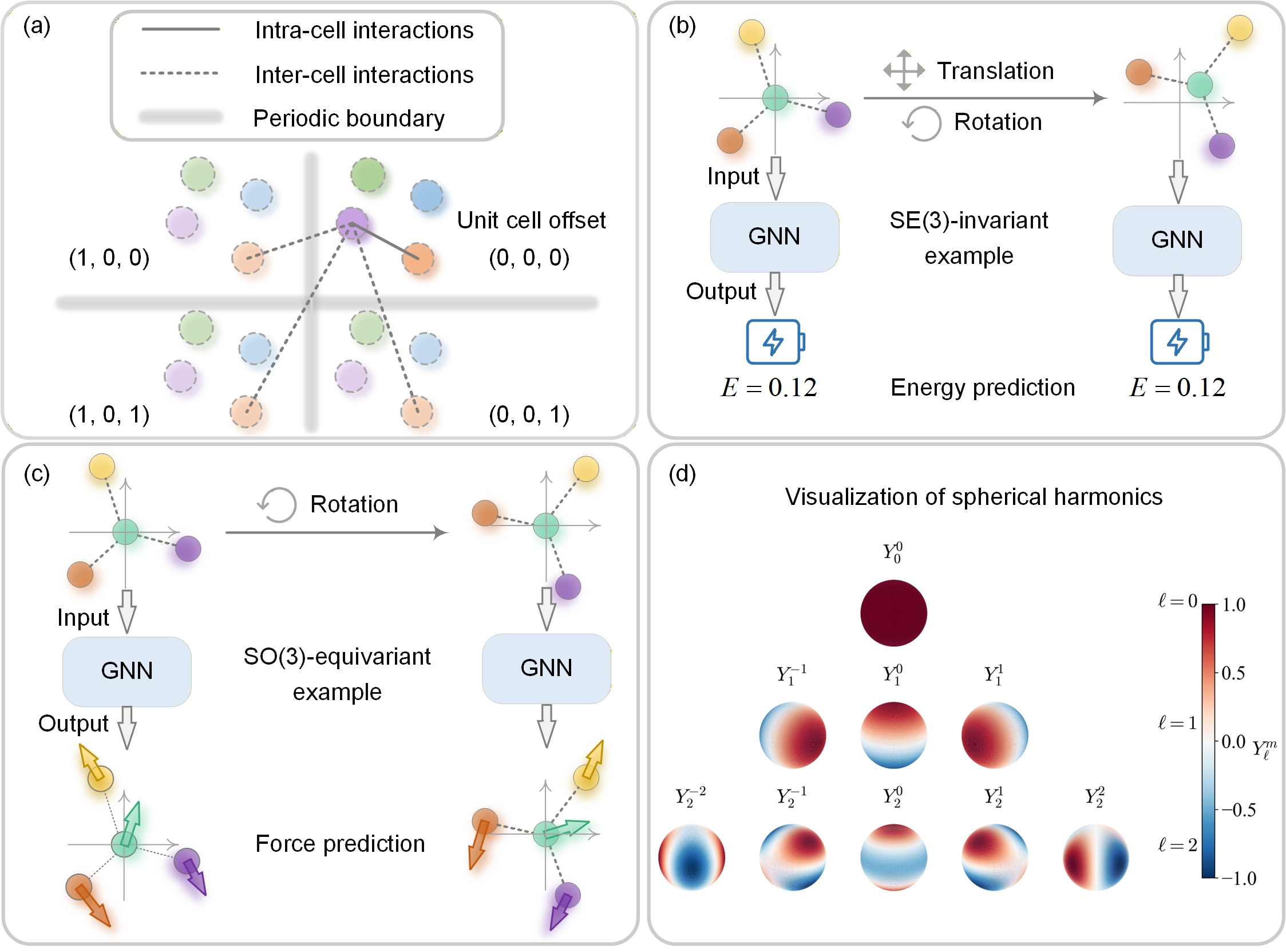}
  \caption{Background concepts for GNN-based crystal structure modeling.
  (a) Periodic crystal graph illustrating an atom connected to symmetry-equivalent neighbors through different translated unit cells.
  (b) Example of a rotation- and translation-invariant GNN: the predicted energy remains unchanged under global rigid transformations of the input structure.
  (c) Example of a rotation-equivariant GNN: predicted atomic forces transform consistently with rotations of the input.
  (d) Real spherical harmonics $Y_{\ell m}(\theta,\phi)$ for degrees $\ell=0,1,2$, visualized on the unit sphere. Each row corresponds to a fixed degree $\ell$ and contains $(2\ell+1)$ basis functions with orders $m\in\{-\ell,\ldots,\ell\}$.}
  \label{fgr:background}
\end{figure*}

\section{Related Work}
\subsection{Iterative ML Models for Structural Optimization}
Iterative ML methods accelerate DFT-based structural optimization by replacing the expensive self-consistent inner loop with a learned surrogate, typically a GNN that predicts energies, forces, and stresses, while still retaining the outer geometry-update loop. A central challenge in this setting is to design GNN architectures that can accurately capture atomic interactions. To ensure physically consistent energy predictions under translations, rotations, and reflections, many earlier state-of-the-art models enforce E(3)-invariance by relying on geometric features such as bond lengths and bond angles \cite{xie2018crystal, schutt2017schnet, deng2023chgnet, gasteiger_dimenet_2020, choudhary2021atomistic, chen2022universal, gasteiger2021gemnet, Omee2024, li2025critical}. More recently, equivariant networks, which explicitly incorporate crystal symmetries and provide richer geometric representations, have demonstrated superior accuracy \cite{batatia2022mace, batzner20223, xu2024equivariant, wang2025elora, park2024scalable, schutt2021equivariant, deng2023chgnet}. 

Despite their success, iterative ML potentials still require extensive supervision from energies, forces, and stresses, and they retain an outer optimization loop. As a result, inference remains inherently sequential, which limits parallel scalability.

\subsection{Iteration-Free Models for Structural Optimization}
Iteration-free methods eliminate both the inner self-consistent loop and the outer geometry-update loop \cite{yang2024scalable, yoon2020differentiable, kim2023structure, yang2025modeling}, and instead aim to map an unrelaxed structure directly to its relaxed counterpart within a single predictive pipeline. 

However, current state-of-the-art approaches, such as DeepRelax \cite{yang2024scalable}, DOGSS \cite{yoon2020differentiable}, and Cryslator \cite{kim2023structure}, typically rely on multi-stage designs and are therefore not fully end-to-end. Moreover, these methods focus primarily on atomic coordinates and do not explicitly represent lattice vectors within the model. Since lattice geometry directly influences material properties such as electronic band structure, density, and mechanical behavior, the lack of explicit lattice modeling may limit predictive accuracy and lead to physically less consistent structural predictions.

\subsection{Expressive Equivariant GNNs for Atomistic Modeling}
Recent equivariant GNNs for atomistic modeling have sought to improve geometric expressivity by capturing finer angular structure and richer many-body interactions. However, prior approaches have struggled to strengthen both aspects simultaneously without incurring substantial computational cost.

Classical SO(3)-equivariant models improve angular expressivity by increasing the degree of spherical harmonic features, but this typically requires Clebsch--Gordan tensor products whose cost scales as $\mathcal{O}(\ell_{\max}^6)$ \cite{batzner20223, batatia2022mace, wen2024equivariant}. Recent works such as HEGNN \cite{cen2024high} and SO3KRATES \cite{frank2024euclidean, frank2022so3krates} reduce this cost by extracting scalar filters from learned high-degree steerable features. Nevertheless, they still maintain and update high-degree equivariant variables $\tilde{\mathbf{h}}_i^{\ell}$ throughout message passing, which can increase memory footprint and architectural complexity.

From a complementary perspective, models such as MACE \cite{batatia2022mace} improve many-body expressivity by explicitly constructing higher-order interactions through iterative tensor products of atomic cluster expansion bases. While theoretically powerful, such constructions become increasingly expensive as the interaction order grows and can be difficult to scale to large systems.

Overall, existing methods still face a trade-off between geometric expressivity and computational efficiency. In particular, many prior approaches either maintain high-degree equivariant features throughout message passing or rely on expensive tensor-product constructions to model higher-order interactions.

\section{Background}
In this section, we introduce the necessary mathematical foundations and conceptual background for the subsequent sections. We denote invariant scalar features by $\mathbf{h}$, equivariant vector features by $\vec{\mathbf{h}}$, and high-degree steerable features by $\tilde{\mathbf{h}}$.

\subsection{Crystal Structures and Periodic Crystal Graphs}
A crystal structure is described by a set of atomic species and positions $\{(z_i, \vec{\mathbf{r}}_i)\}_{i=1}^N$, with $\vec{\mathbf{r}}_i \in \mathbb{R}^3$, together with a lattice matrix $\mathbf{L} = [\vec{\bm{l}}_1, \vec{\bm{l}}_2, \vec{\bm{l}}_3] \in \mathbb{R}^{3\times 3}$ whose columns are the primitive lattice vectors. Under periodic boundary conditions, the lattice generates an infinite set of symmetry-equivalent images, i.e., any atom at position \(\vec{\mathbf{r}}\) is replicated at \(\vec{\mathbf{r}} + \mathbf{L}\mathbf{k}\), where \(\mathbf{k} = (k_1,k_2,k_3) \in \mathbb{Z}^3\) is the unit cell offset. This induces a periodic crystal graph \(\mathcal{G} = (\mathcal{V}, \mathcal{E})\), where \(\mathcal{V}\) denotes the set of atomic nodes and \(\mathcal{E}\) the set of edges representing pairwise interactions.  Each edge \((i, j, \mathbf{k}) \in \mathcal{E}\) specifies that the interaction between atoms $i$ and $j$ occurs through the periodic image of atom $j$ translated by \(k_1 \vec{\bm{l}}_1 + k_2 \vec{\bm{l}}_2 + k_3 \vec{\bm{l}}_3\). Consequently, a given atom $i$ may connect to the same atom $j$ through multiple edges, each corresponding to a different periodic image specified by its own $\mathbf{k}$ (Figure \ref{fgr:background} (a)). The graph connectivity is typically determined using a distance cut-off $D$ together with a maximum number of neighbors $K$.

\subsection{Graph Neural Networks}
Graph neural networks (GNNs) provide a natural framework for modeling periodic crystal graphs. Their core computational mechanism is message passing \cite{gilmer2017neural, han2025survey}, in which each node aggregates information from its neighbors to update its hidden representation. Let \(\mathbf{h}_i^{(t)}\) denote the hidden feature of node \(i\) at iteration \(t\), and let \(\vec{\mathbf{r}}_{ji} = \vec{\mathbf{r}}_i - \vec{\mathbf{r}}_j\) denote the relative position vector (for simplicity we omit periodic offsets $\mathbf{k}$). A typical message-passing layer takes the form
\begin{equation}
\mathbf{m}_{ji} = M_t\!\left(\mathbf{h}_i^{(t)},\mathbf{h}_j^{(t)}, \Vert \vec{\mathbf{r}}_{ji} \Vert \right),
\label{eq:mp_message}
\end{equation}
\begin{equation}
\mathbf{h}_i^{(t+1)} = U_t\!\left(\mathbf{h}_i^{(t)},\,\sum_{j \in \mathcal{N}(i)}\mathbf{m}_{ji} \right),
\label{eq:mp_update}
\end{equation}
where $\mathcal{N}(i)$ denotes the neighbors of node $i$, and $M_t$ and
$U_t$ are learnable differentiable functions.

\subsection{Symmetry in Physics}
Respecting the symmetries of the underlying physical system is crucial for designing neural networks for atomistic modeling. Scalar quantities such as total energy must remain invariant under global translations and rotations, whereas vector and tensor quantities (e.g., forces, stresses) must transform equivariantly under the same transformations (Figure \ref{fgr:background} (b) and (c)).

Let $\{T_g : \mathcal{X} \to \mathcal{X}\}_{g\in G}$ be a group action of an abstract group $G$ on the input space $\mathcal{X}$. A function $\phi : \mathcal{X} \to \mathcal{Y}$ is said to be \emph{equivariant} to this action if there exists a corresponding group action $\{S_g : \mathcal{Y} \to \mathcal{Y}\}_{g\in G}$ on the output space such that:
\begin{equation}
    \phi(T_g(x)) = S_g(\phi(x)), \qquad \forall g\in G,\, x\in\mathcal{X}.
\end{equation}

If the output of $\phi$ is a scalar, i.e.\ $\phi(x)\in\mathbb{R}$, then the only valid group action on $\mathbb{R}$ is the identity map. In this case, $\phi$ is \emph{invariant} to the group action:
\begin{equation}
    \phi(T_g(x)) = \phi(x).
\end{equation}
By definition, invariance is therefore a special case of equivariance in which the output transforms trivially (i.e., remains unchanged) under the group action.

\subsection{Spherical Harmonics}
\label{sec:sh_rep}
The relative position vector $\vec{\mathbf{r}}_{ji} = \vec{\mathbf{r}}_i - \vec{\mathbf{r}}_j$ can be decomposed into radial and angular components:
\begin{equation}
    \vec{\mathbf{r}}_{ji} = \|\vec{\mathbf{r}}_{ji}\|\,\hat{\mathbf{r}}_{ji},
\end{equation}
where $\hat{\mathbf{r}}_{ji}$ is the unit direction vector. Since $\hat{\mathbf{r}}_{ji}$ can be parameterized by spherical coordinates $(\theta,\phi)$, any function dependent on the bond orientation $j\!\to\!i$ essentially models an angular dependence on $(\theta,\phi)$.

To represent these angular dependencies under the action of SO(3), it is standard to expand them using spherical harmonics. The real spherical harmonics $Y_{\ell m}(\hat{\mathbf{r}})$ are indexed by a degree $\ell \ge 0$, representing the angular frequency, and an order $m \in \{-\ell,\dots,\ell\}$. For a fixed degree $\ell$, the set
\begin{equation}
    \bigl\{\, Y_{\ell m}(\hat{\mathbf{r}}_{ji}) \;\big|\; m=-\ell,\dots,\ell \,\bigr\}
\end{equation}
forms a $(2\ell+1)$-dimensional basis that captures all angular information of frequency $\ell$ on the unit sphere (Figure \ref{fgr:background} (d)).

A key property of spherical harmonics is that they transform under rotations according to the Wigner-$D$ representation. For any rotation $g \in \mathrm{SO}(3)$,
\begin{equation}
    Y_{\ell m}(g \cdot \hat{\mathbf{r}}_{ji})
    \;=\;
    \sum_{m'=-\ell}^{\ell}
    D^{\ell}_{m m'}(g)\,
    Y_{\ell m'}(\hat{\mathbf{r}}_{ji}),
\end{equation}
where $D^\ell(g)$ is the $(2\ell+1)\times(2\ell+1)$ Wigner matrix for degree~$\ell$.

\subsection{Invariant and Equivariant GNNs}
\label{sec:gnns}
Graph neural networks for atomistic modeling can be broadly categorized based on how they handle geometric symmetries, i.e., invariant and equivariant GNNs.

\subsubsection{Invariant GNNs} 
A common approach for respecting SE(3) symmetry is to build features from geometric quantities that are intrinsically invariant under the SE(3) symmetry group, such as interatomic distances, bond angles, and dihedral angles, yielding what are known as invariant GNNs \cite{xie2018crystal, deng2023chgnet, chen2022universal, choudhary2021atomistic, gasteiger_dimenet_2020, gasteiger2021gemnet, liu2022spherical, yan2022periodic, lin2023efficient, yang2024interaction}.

Models relying solely on interatomic distances are limited to capturing two-body interactions (i.e., pairwise geometric relations between an atom $i$ and neighbor $j$). To resolve structural ambiguities and enhance expressiveness, many architectures incorporate bond angles and dihedrals. These features introduce many-body interactions, e.g., triplets and quadruplets of atoms, respectively, thereby providing a more complete description of the local atomic environment.

While invariant GNNs are generally computationally efficient, they have a limitation, i.e., they discard orientation information. This loss fundamentally hinders the network’s ability to propagate directional cues effectively across layers.

\subsubsection{Equivariant GNNs} 
Equivariant models overcome the limitations of invariant architectures by carrying vector and tensor features that rotate coherently with the underlying geometry, thus able to represent orientation fields and propagate directional dependencies across layers \cite{fuchs2020se, brandstetter2021geometric, satorras2021n, shuaibi2021rotation, batzner2023advancing, musaelian2023learning, batzner20223, passaro2023reducing, zitnick_scn_2022, batatia2022mace, equiformer_v2, aykent2025gotennet, cen2024high, du2022se, frank2024euclidean, wen2024equivariant, yin2025alphanet, yang2024interaction, gong2023general, wen2025cartesian, batatia2025design, 11098662}.

In equivariant GNNs such as Tensor Field Networks \cite{thomas2018tensor}, message passing relies on Clebsch-Gordan tensor products that couple feature tensors with geometric information. Given node features of degree $\ell_i$ and spherical harmonics of degree $\ell_f$, the output message of degree~$\ell_o$ is computed as
\begin{equation}
    \tilde{\mathbf{m}}_{ji}^{\ell_o m_o} 
    =
    \sum_{\ell_i, \ell_f, m_i, m_f}
    C^{\ell_o m_o}_{\ell_i m_i \ell_f m_f}\,
    \phi^{\,\ell_i \ell_f}_{\ell_o}(\Vert \vec{\mathbf{r}}_{ji} \Vert)\,
    Y_{\ell_f m_f}(\hat{\mathbf{r}}_{ji})\,
    \tilde{\mathbf{h}}^{\,\ell_i m_i}_{j}
\end{equation}
where $C^{\ell_o m_o}_{\ell_i m_i\ell_f m_f}$ are the Clebsch-Gordan coefficients and $\phi^{\ell_i \ell_f}_{\ell_o}$ is a learnable radial kernel.

While SO(3)-equivariant convolutions are highly expressive, their tensor-product operations are computationally expensive. The naive complexity grows as $\mathcal{O}(\ell_{\max}^6)$, and practical implementations can be up to two orders of magnitude slower than invariant GNNs \cite{fu2023forces}.

\subsection{Scalar-Vector Representations}
A common strategy to reduce the computational cost of equivariant GNNs is to restrict the model to the lowest nontrivial irreducible representations of $\mathrm{SO}(3)$, namely, the scalar ($\ell=0$) and vector ($\ell=1$) components \cite{schutt2021equivariant, haghighatlari2022newtonnet, yang2025efficient, satorras2021n}. In this scalar-vector representation scheme, each node feature is decomposed into an $\ell=0$ invariant scalar channel and an $\ell=1$ equivariant vector channel:
\begin{equation}
    \mathbf{H}_i = 
    \bigl(
        \mathbf{h}_i,\;
        \vec{\mathbf{h}}_i
    \bigr),
\end{equation}
where $\mathbf{h}_i \in \mathbb{R}^{C}$ is rotation-invariant and $\vec{\mathbf{h}}_i \in \mathbb{R}^{C \times 3}$ transforms as a 3D vector under any rotation.

Restricting to $\ell \le 1$ substantially lowers computational complexity by avoiding high-dimensional tensor products. However, this simplification also limits expressive power, i.e., typical scalar-vector models cannot represent higher-degree angular frequencies ($\ell\!\ge\!2$) and therefore struggle to capture fine-grained geometric distinctions \cite{cen2024high}.

\subsection{Geometric Expressivity: Degree and Order}
\label{sec:expressivity}
In geometric deep learning, expressivity refers to a model's capacity to distinguish between geometrically distinct atomic environments that might otherwise appear identical under simpler representations. In this work, we distinguish two complementary axes of geometric expressivity, i.e., degree (angular resolution) and order (many-body correlations).

\subsubsection{Degree}
The degree of a representation refers to the maximum angular frequency $\ell_{\text{max}}$ used in its spherical-harmonic expansion. As detailed in Section~\ref{sec:sh_rep}, a feature of degree $\ell$ transforms according to a $(2\ell+1)$-dimensional irreducible representation of $\mathrm{SO}(3)$. Models restricted to $\ell=0$ (scalars) or $\ell=1$ (vectors) have limited angular resolution \cite{cen2024high}. We refer to high-degree expressivity as the use of features with $\ell \ge 2$, which allow the network to resolve sharper angular variations and more anisotropic interactions.

\subsubsection{Order}
The order of an interaction refers to the number of atoms that are coupled simultaneously in a single message-passing step. A standard distance-based message-passing layer is inherently two-body, depending only on pairs $(i,j)$. Higher-order expressivity arises when the interaction depends on triplets $(i,j,k)$, quadruplets $(i,j,k,l)$, or larger clusters. Physically, such terms capture bond angles, torsion angles, and collective local distortions that are essential for modeling realistic crystal potential energy surfaces. Unlike degree, which controls angular resolution, increasing body order captures qualitatively richer many-body correlations.

Crucially, these two axes are distinct, i.e., a model can be high-degree but only two-body (e.g., using high-$\ell$ tensor products on a single edge), or purely scalar but high-order (e.g., using angle- and dihedral-based invariants).

\begin{figure*}[t]
  \centering
  \includegraphics[width=1.8\columnwidth]{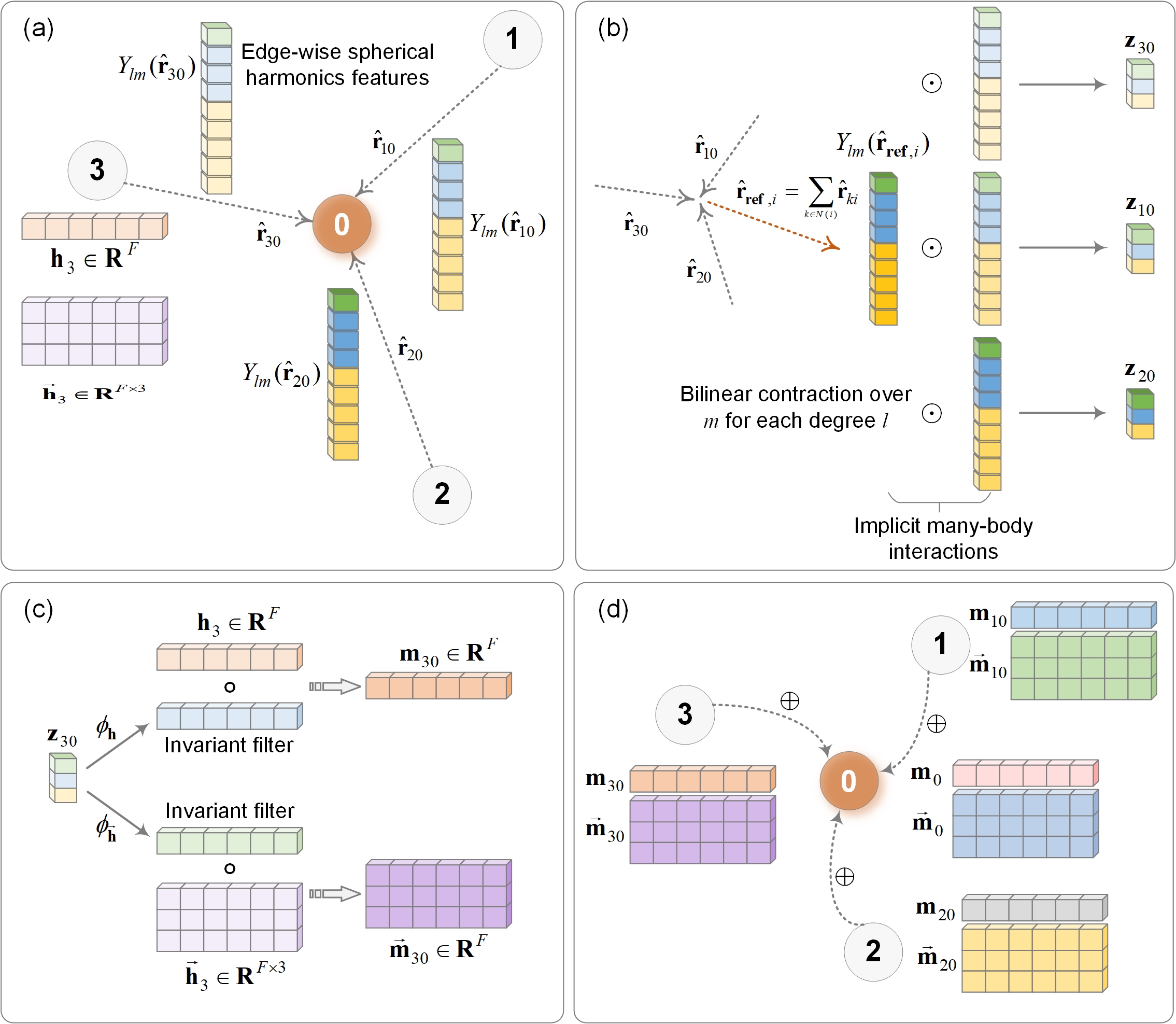}
  \caption{Overview of the proposed high-degree, high-order message passing (H$^2$-MP).
(a) For a central atom $i$ (node $0$), spherical harmonics $Y_{\ell m}(\hat{\mathbf{r}}_{ji})$ up to degree $\ell_{\max}$ are evaluated on the directions from neighboring atoms $j \in \mathcal{N}(i)$, providing high-degree angular descriptors for each edge.
(b) A local reference direction $\hat{\mathbf{r}}_{\mathrm{ref},i}$ is constructed by aggregating all neighbor directions, and degree-wise bilinear invariants $z_{ji}^\ell$ are obtained by contracting spherical harmonics of $\hat{\mathbf{r}}_{ji}$ and $\hat{\mathbf{r}}_{\mathrm{ref},i}$. 
(c) The geometric invariants $\mathbf{z}_{ji}$ are mapped to invariant filters that gate scalar and vector edge messages without maintaining persistent high-degree equivariant features.
(d) Edge-wise scalar and vector messages are aggregated over neighbors to update node features.}
  \label{fgr:E3RelaxH2}
\end{figure*}

\section{Methodology}
E$^{3}$Relax-H$^{2}$ is designed to address three key challenges in iteration-free crystal structure optimization: (i) achieving strong geometric expressivity without prohibitive computational cost, (ii) modeling the coupled evolution of atomic coordinates and lattice geometry in a symmetry-consistent manner, and (iii) providing physically meaningful supervision under periodic boundary conditions. To tackle these challenges, the proposed framework integrates three core components: (i) a high-degree, high-order message-passing module (H$^2$-MP), which enhances geometric expressivity while maintaining the efficiency of scalar–vector architectures; (ii) a lattice–atom message-passing module (LA-MP), which explicitly captures the bidirectional coupling between atomic displacements and lattice deformation; and (iii) a differentiable periodicity-aware Cartesian displacement loss, which provides physically consistent supervision for one-shot prediction in periodic systems. 

At a high level, E$^{3}$Relax-H$^2$ takes an initial periodic crystal $(\{z_i, \vec{\mathbf{r}}_i\}_{i=1}^{N}, \mathbf{L})$ as input and processes it through a stack of equivariant interaction layers. Each layer contains H$^2$-MP and self-interaction, together with an interleaved LA-MP module comprising lattice-to-atom broadcast, atom-to-lattice aggregation, and lattice-lattice interaction, followed by structure update. Within each layer, the computation proceeds in the following order: lattice-to-atom broadcast, H$^2$-MP, self-interaction, atom-to-lattice aggregation, lattice-lattice interaction, and structure update. Stacking $T$ such layers produces a sequence of intermediate structures and ultimately a relaxed crystal configuration. 

\subsection{Input Encoding}
We represent a crystal as a dual-node graph \(\mathcal{G} = (\mathcal{V}^{a}, \mathcal{V}^{l}, \mathcal{E})\), where \(\mathcal{V}^{a} = \{1,\dots,N\}\) indexes atom nodes and \(\mathcal{V}^{l} = \{1,2,3\}\) indexes lattice nodes corresponding to the three lattice basis vectors. This dual-node formulation makes the lattice an explicit part of the graph structure. 

Each atom node \(i \in \mathcal{V}^{a}\) carries an invariant scalar feature \(\mathbf{h}_i \in \mathbb{R}^{F}\), an equivariant vector feature \(\vec{\mathbf{h}}_i \in \mathbb{R}^{F \times 3}\), and a Cartesian position \(\vec{\mathbf{r}}_i \in \mathbb{R}^{3}\). Each lattice node \(c \in \mathcal{V}^{l}\) is assigned analogous features
\(\mathbf{s}_c \in \mathbb{R}^{F}\), \(\vec{\mathbf{s}}_c \in \mathbb{R}^{F \times 3}\), and the lattice vector \(\vec{\bm{l}}_c \in \mathbb{R}^{3}\). Initialization proceeds as follows. 

The atomic scalar feature \(\mathbf{h}_i^{(0)}\) is obtained from the atomic number via an embedding layer \(E_{\text{atom}}\colon \mathbb{Z} \to \mathbb{R}^{F}\), and the atomic vector feature is initialized to zero, \(\vec{\mathbf{h}}_i^{(0)} = \vec{\mathbf{0}} \in \mathbb{R}^{F \times 3}\). The atomic coordinates \(\vec{\mathbf{r}}_i^{(0)}\) are set to the positions of the initial (unrelaxed) structure.

For lattice nodes, the scalar feature \(\mathbf{s}_c^{(0)}\) is initialized from a learnable embedding \(E_{\text{lat}}\colon \{1,2,3\} \to \mathbb{R}^{F}\), and the lattice vector feature is initialized to zero, \(\vec{\mathbf{s}}_c^{(0)} = \vec{\mathbf{0}} \in \mathbb{R}^{F \times 3}\). The coordinates of lattice node \(\vec{\bm{l}}_c^{(0)}\) are initialized to the corresponding lattice vectors of the initial unrelaxed unit cell.

\subsection{High-Degree, High-Order Message Passing}
\label{sec:h2mp}
To address the expressivity-efficiency challenge, we introduce a high-degree, high-order message passing module (H$^2$-MP). The H$^2$-MP module computes geometry-aware scalar and vector messages, gated by high-degree, many-body-aware angular invariants $z^\ell_{ji}$.

We construct $z_{ji}^{\ell}$ from purely geometric information. For each directed edge $(j \to i)$, let
$\hat{\mathbf{r}}_{ji}$ be the unit direction from atom $j$ to atom $i$. We define a local reference direction at node $i$ by averaging neighbor directions:
\begin{equation}
    \hat{\mathbf{r}}_{\mathrm{ref}, i}
    \;\propto\;
    \sum_{k \in \mathcal{N}(i)} \hat{\mathbf{r}}_{ki},
    \qquad
    \|\hat{\mathbf{r}}_{\mathrm{ref}, i}\| = 1,
\end{equation}
and evaluate spherical harmonics up to degree $\ell_{\max}$ on both $\hat{\mathbf{r}}_{ji}$ and $\hat{\mathbf{r}}_{\mathrm{ref}, i}$. A bilinear map then produces, for each degree $\ell$, a scalar invariant $z_{ji}^{\ell}$ that measures the angular alignment between the edge direction and the local reference frame:
\begin{equation}
    z_{ji}^{\ell}
    =
    \sum_{m=-\ell}^{\ell}
    Y_{\ell m}(\hat{\mathbf{r}}_{ji})\,
    Y_{\ell m}(\hat{\mathbf{r}}_{\mathrm{ref},i}),
\end{equation}

These invariants $z_{ji}^{\ell}$ provide high-degree angular resolution without maintaining persistent high-degree feature channels $\tilde{\mathbf{h}}_i^{\ell}$, in contrast to HEGNN and SO3KRATES. Moreover, although $z_{ji}^{\ell}$ is indexed by a single edge $(j,i)$, it depends on $\hat{\mathbf{r}}_{\mathrm{ref}, i}$, which in turn is a function of all neighbor directions $\{\hat{\mathbf{r}}_{ki}\}_{k \in \mathcal{N}(i)}$. Each $z_{ji}^{\ell}$ therefore couples atom $j$ with atom $i$ and the entire neighborhood of $i$, implicitly encoding many-body correlations around $i$.

We collect the invariants for all degrees into an edge-wise geometric descriptor
\begin{equation}
    \mathbf{z}_{ji}
    =
    \bigoplus_{\ell=0}^{\ell_{\max}} z_{ji}^{\ell},
\end{equation}
where $\bigoplus$ denotes concatenation. A small MLP $\phi_{\mathrm{gate}}$ then maps $\mathbf{z}_{ji}$ to a geometry-dependent gate,
\begin{equation}
    \mathbf{w}_{ji}
    =
    \tanh\!\bigl(
        \phi_{\mathrm{gate}}(\mathbf{z}_{ji})
    \bigr)
    \;\in\; \mathbb{R}^{F},
\end{equation}
which is used to modulate the strength and angular profile of the messages later.

Given radial edge features
${\mathbf{e}}_{ji} = \gamma_h\bigl(\lambda(\|\vec{\mathbf{r}}_{ji}\|)\bigr)$
obtained from a radial basis expansion $\lambda(\cdot)$ followed by an
MLP $\gamma_h(\cdot)$, and a node-wise projection
$\phi_{\mathbf{h}}(\mathbf{h}_j^{(t)})$, we define edge-wise invariant
messages
\begin{equation}
    \mathbf{m}_{ji}
    =
    \mathbf{w}_{ji}
    \;\circ\;
    \phi_{\mathbf{h}}(\mathbf{h}_j^{(t)})
    \;\circ\;
    {\mathbf{e}}_{ji}^{(t)},
\end{equation}
where $\circ$ denotes element-wise multiplication.

The node-level scalar and vector messages at node $i$ are then computed as
\begin{equation}
    \mathbf{m}_i
    =
    \sum_{j \in \mathcal{N}(i)} \phi_{\mathbf{m}}(\mathbf{m}_{ji}),
\end{equation}
\begin{equation}
    \vec{\mathbf{m}}_i
    =
    \sum_{j \in \mathcal{N}(i)}
    \Bigl[
        \phi_{\vec{\mathbf{h}}}(\mathbf{m}_{ji}) \;\circ\; \vec{\mathbf{h}}_j^{(t)}
        \;+\;
        \phi_{\vec{\mathbf{r}}}(\mathbf{m}_{ji}) \;\circ\; \hat{\mathbf{r}}_{ji}^{(t)}
    \Bigr],
\end{equation}
where $\phi_{\mathbf{m}}$, $\phi_{\vec{\mathbf{h}}}$, and $\phi_{\vec{\mathbf{r}}}$ are small invariant MLPs. These intermediate messages $\mathbf{m}_i$ and $\vec{\mathbf{m}}_i$ are subsequently used in the self-interaction module to update $\mathbf{h}_i^{(t)}$ and $\vec{\mathbf{h}}_i^{(t)}$.

\subsection{Self-Interaction}
The self-interaction module refines each node’s features by fusing its scalar and vector messages. Given the aggregated scalar and vector messages $\mathbf{m}_i$ and $\vec{\mathbf{m}}_i$, we first form a rotation-invariant descriptor by contracting the vector message with a learnable linear map $\mathbf{U} \in \mathbb{R}^{F \times F}$ and taking its norm. The scalar and vector updates are then computed as
\begin{equation}
    \mathbf{h}_i^{(t+1)}
    =
    \phi_{s1}\!\bigl(\mathbf{m}_{i} \oplus \|\mathbf{U}\,\vec{\mathbf{m}}_i\|\bigr)
    \;+\;
    \tanh\!\bigl(
        \phi_{s2}\!\bigl(\mathbf{m}_{i} \oplus \|\mathbf{U}\,\vec{\mathbf{m}}_i\|\bigr)
    \bigr)\,\mathbf{m}_i,
\end{equation}
\begin{equation}
    \vec{\mathbf{h}}_i^{(t+1)}
    =
    \phi_{v}\!\bigl(\mathbf{m}_{i} \oplus \|\mathbf{U}\,\vec{\mathbf{m}}_i\|\bigr)
    \;\circ\;
    \bigl(\mathbf{V}\,\vec{\mathbf{m}}_i\bigr),
\end{equation}
where $\oplus$ denotes concatenation, $\phi_{s1}, \phi_{s2}, \phi_{v} : \mathbb{R}^{2F} \rightarrow \mathbb{R}^F$ are MLPs, $\mathbf{V} \in \mathbb{R}^{F \times F}$ is a learnable matrix, and $\|\mathbf{U}\,\vec{\mathbf{m}}_i\|$ denotes the channel-wise norm taken over the Cartesian dimension. The $\tanh(\cdot)$ term acts as a scalar gate that controls how much of the incoming scalar message $\mathbf{m}_i$ is preserved in $\mathbf{h}_i^{(t+1)}$.

\subsection{Lattice-Atom Message Passing}
\label{sec:lamp}
To explicitly model the coupled evolution of atomic coordinates and lattice geometry, we introduce the LA-MP module that models the joint evolution of atomic coordinates and lattice vectors under SE(3)-equivariance. 

Concretely, we represent the three lattice basis vectors as explicit lattice nodes and connect each lattice node to all atoms in the unit cell, enabling bidirectional information flow between atomic and lattice states. Each lattice node maintains scalar and vector states $\mathbf{s}_{c}\in\mathbb{R}^{F}$ and $\vec{\mathbf{s}}_{c}\in\mathbb{R}^{F \times 3}$ for $c\in\mathcal{V}^l$, and updates them together with atomic features through three stages: (i) lattice-to-atom broadcast, (ii) atom-to-lattice aggregation, and (iii) lattice-lattice interaction. 

Importantly, the two directions of lattice-atom coupling are realized at different points within each interaction layer: lattice-to-atom broadcast is performed before atomic message passing, whereas atom-to-lattice aggregation is carried out after the self-interaction step.

\subsubsection{Lattice-to-Atom Broadcast}
The lattice-to-atom broadcast stage injects global lattice information into local atomic states, allowing atomic updates to be explicitly conditioned on the current lattice geometry. 

Compared with E$^{3}$Relax, the present design restores translation invariance in lattice-atom coupling by constructing the interaction from centered atomic coordinates and translation-consistent geometric descriptors. Specifically, to ensure translation invariance, we first define centered atomic coordinates
\begin{equation}
\tilde{\vec{\mathbf{r}}}_i = \vec{\mathbf{r}}_i - \bar{\vec{\mathbf{r}}},
\end{equation}
where $\bar{\vec{\mathbf{r}}}$ denotes the mean atomic position of the structure. Using the centered atomic coordinate $\tilde{\vec{\mathbf{r}}}_i$ and lattice basis vector $\vec{\bm{l}}_c$, we characterize the geometric relationship between atom $i$ and lattice axis $c$ through two complementary quantities, i.e., the parallel projection and the perpendicular distance relative to the lattice axis. The parallel component is defined as
\begin{equation}
d_{ic}^{\parallel}
=
\frac{\tilde{\vec{\mathbf{r}}}_i \cdot \vec{\bm{l}}_c}
{\|\vec{\bm{l}}_c\|},
\end{equation}
while the perpendicular distance is given by
\begin{equation}
d_{ic}^{\perp}
=
\frac{\|\tilde{\vec{\mathbf{r}}}_i \times \vec{\bm{l}}_c\|}
{\|\vec{\bm{l}}_c\|}.
\end{equation}
These quantities measure the projection of the atomic displacement along the lattice axis and the distance to that axis, respectively. Based on these geometric descriptors, we construct an invariant lattice-atom encoding $\mathbf{e}_{ci}$ by applying radial basis expansions to $d_{ic}^{\parallel}$ and $d_{ic}^{\perp}$, concatenating the resulting features, and processing them with an MLP.

To propagate equivariant information, we further define the lattice-atom direction
\begin{equation}
\hat{\mathbf{r}}_{ci}
=
\frac{\vec{\bm{l}}_c - \tilde{\vec{\mathbf{r}}}_i}
{\|\vec{\bm{l}}_c - \tilde{\vec{\mathbf{r}}}_i\|},
\end{equation}
which serves as the geometric carrier for vector messages.

Given the invariant encoding $\mathbf{e}_{ci}$ and direction $\hat{\mathbf{r}}_{ci}$, lattice information is broadcast to atomic states through gated message updates.

The scalar atomic features are updated as
\begin{equation}
\mathbf{h}_i^{+}
=
\sum_{c=1}^{3}
\Bigl[
\phi_{s}^{a}(\mathbf{h}_i^{(t)})
\circ
\mathbf{e}_{ci}^{(t)}
+
\phi_{s}^{l}(\mathbf{s}_c^{(t)})
\Bigr]
+
\mathbf{h}_i^{(t)},
\label{eq:la_s}
\end{equation}
where $\phi_{s}^{a}$ and $\phi_{s}^{l}$ are MLPs.

The vector atomic features are updated as
\begin{equation}
    \vec{\mathbf{h}}_i^{+}
=
\phi_{v}^{a}\!\bigl(\mathbf{h}_i^{+}\bigr)\circ \vec{\mathbf{h}}_i^{(t)}
+
\sum_{c=1}^{3}
\Bigl[
\phi_{v}^{r}\!\bigl(\mathbf{h}_i^{+}\bigr)\circ \hat{\mathbf{r}}_{ci}^{(t)}
+
\mathbf{W}_c\,\vec{\mathbf{s}}_c^{(t)}
\Bigr]
+
\vec{\mathbf{h}}_i^{(t)},
\label{eq:la_v}
\end{equation}
where $\phi_{v}^{a}$ and $\phi_{v}^{r}$ are invariant MLPs, and $\mathbf{W}_c \in \mathbb{R}^{F \times F}$ is a learnable linear projection. 

The updated atomic features $\mathbf{h}_i^{+}$ and $\vec{\mathbf{h}}_i^{+}$ are then passed to the subsequent H$^2$-MP and self-interaction modules. The refined atomic features are thereafter aggregated back to the lattice nodes through the atom-to-lattice aggregation stage described below.

\subsubsection{Atom-to-Lattice Aggregation}
After H$^2$-MP and self-interaction, atomic information is aggregated back to the lattice nodes to update the lattice states.

Specifically, we first aggregate atomic features over all atoms in the unit cell using mean pooling:
\begin{equation}
\bar{\mathbf{h}}
=
\frac{1}{|\mathcal{V}^a|}
\sum_{i \in \mathcal{V}^a} \mathbf{h}_i^{(t+1)},
\end{equation}
\begin{equation}
\bar{\vec{\mathbf{h}}}
=
\frac{1}{|\mathcal{V}^a|}
\sum_{i \in \mathcal{V}^a} \vec{\mathbf{h}}_i^{(t+1)} ,
\end{equation}
Then, for each lattice node $c \in \mathcal{V}^l$, the pooled atomic features are fused with the current lattice state to update scalar and vector lattice features:
\begin{align}
\mathbf{s}_c^{+} &=
\mathbf{s}_c^{(t)}
+
\phi_{a\rightarrow l}^{s}
\!\left(
\bar{\mathbf{h}} \oplus \mathbf{s}_c^{(t)}
\right), \\
\vec{\mathbf{s}}_c^{+} &=
\vec{\mathbf{s}}_c^{(t)}
+
\phi_{a\rightarrow l}^{v}
\!\left(
\bar{\vec{\mathbf{h}}} + \vec{\mathbf{s}}_c^{(t)}
\right),
\end{align}
where $\phi_{a\rightarrow l}^{s}$ is an invariant MLP and $\phi_{a\rightarrow l}^{v}$ is a linear equivariant map applied channel-wise.

\subsubsection{Lattice-Lattice Interaction}
In addition to lattice-atom coupling, interactions among lattice basis vectors are essential for modeling coherent cell deformations. To capture these effects, we introduce a lattice-lattice interaction stage that enables information exchange among the three lattice vectors.

We first stack the per-axis lattice states as
\begin{equation}
    \mathbf{S}^+ = [\mathbf{s}_1^+, \mathbf{s}_2^+, \mathbf{s}_3^+]^\top \in \mathbb{R}^{3 \times F},
\end{equation}
\begin{equation}
    \vec{\mathbf{S}}^+ = [\vec{\mathbf{s}}_1^+, \vec{\mathbf{s}}_2^+, \vec{\mathbf{s}}_3^+]^\top \in \mathbb{R}^{3 \times F \times 3}.
\end{equation}
Lattice-lattice interaction is then realized by mixing lattice states along the axis dimension using shared linear transformations:
\begin{align}
\mathbf{S}^{(t+1)} &= \mathbf{W}_{s}\,\mathbf{S}^+, \\
\vec{\mathbf{S}}^{(t+1)} &= \mathbf{W}_{v}\,\vec{\mathbf{S}}^+,
\end{align}
where $\mathbf{W}_{s} \in \mathbb{R}^{3 \times 3}$ and $\mathbf{W}_{v} \in \mathbb{R}^{3 \times 3}$ are learnable weight matrices that couple the three lattice vectors.

\subsection{Structure Updating}
The structure updating module performs simultaneous updates of (i) atomic Cartesian coordinates and (ii) lattice vectors after each interaction layer. 

After H$^2$-MP, self-interaction, and LA-MP, atom and lattice nodes carry equivariant vector features $\vec{\mathbf{h}}_i^{(t+1)}$ and $\vec{\mathbf{s}}_c^{(t+1)} \in \mathbb{R}^{F \times 3}$, respectively. These vector embeddings are mapped to geometric displacements via linear projections over the feature dimension:
\begin{equation}
    \Delta \vec{\mathbf{r}}_i = \mathbf{W}_{p}\,\vec{\mathbf{h}}_i^{(t+1)},
\end{equation}
\begin{equation}
    \Delta \vec{\bm{l}}_c = \mathbf{W}_{l}\,\vec{\mathbf{s}}_c^{(t+1)},
\end{equation}
where $\mathbf{W}_p,\mathbf{W}_l\in\mathbb{R}^{1 \times F}$ are learnable bias-free linear maps.

Atomic positions and lattice vectors are then updated residually:
\begin{equation}
    \vec{\mathbf{r}}_i^{(t+1)} = \vec{\mathbf{r}}_i^{(t)} + \Delta \vec{\mathbf{r}}_i,
\end{equation}
\begin{equation}
    \vec{\bm{l}}_c^{(t+1)} = \vec{\bm{l}}_c^{(t)} + \Delta \vec{\bm{l}}_c.
\end{equation}

\subsection{Differentiable Periodicity-Aware Cartesian Displacement Loss}
\label{sec:pacd}
To resolve the ambiguity of Cartesian supervision under periodic boundary conditions, we align each prediction to its nearest periodic representative with respect to the ground-truth relaxed structure. This treats boundary-crossing atoms as physically equivalent.

Let $\vec{\mathbf{r}}_{i}^{\text{pred}} \in \mathbb{R}^{3}$ and $\vec{\mathbf{r}}_{i}^{\text{gt}} \in \mathbb{R}^{3}$ denote the predicted and ground-truth relaxed positions of atom $i$. We enumerate the 27 neighbor-cell offsets $\mathcal{K}=\{-1,0,1\}^3$ and choose
\begin{equation}
\mathbf{k}_i^\star
=
\arg\min_{\mathbf{k}\in\mathcal{K}}
\left\|
\vec{\mathbf{r}}_{i}^{\text{pred}} + \mathbf{L}^{\text{gt}}\mathbf{k}
-
\vec{\mathbf{r}}_{i}^{\text{gt}}
\right\|_2,
\end{equation}
The aligned prediction is then given by
\begin{equation}
    \vec{\mathbf{r}}_{i}^{\text{align}}
=
\vec{\mathbf{r}}_{i}^{\text{pred}} + \mathbf{L}^{\text{gt}}\mathbf{k}_i^\star .
\end{equation}
We define the periodicity-aware Cartesian displacement loss as
\begin{equation}
\mathcal{L}_{\text{pos}}
=
\frac{1}{3 \vert \mathcal{V}^a \vert}
\sum_{i \in \mathcal{V}^a}
\left\|
\vec{\mathbf{r}}_{i}^{\text{align}}
-
\vec{\mathbf{r}}_{i}^{\text{gt}}
\right\|_{1}.
\end{equation}

This nearest-image alignment is piecewise differentiable almost everywhere with respect to $\vec{\mathbf{r}}_{i}^{\mathrm{pred}}$. Once the optimal offset is selected, gradients propagate through the aligned coordinate.

\subsection{Layer-Wise Supervision}
If supervision is applied only to the final output, the network may simply copy its intermediate representations forward and perform one large correction in the last layer. Such abrupt “jumps’’ are hard to learn. Instead, E$^{3}$Relax-H$^2$ attaches a loss to every layer so that each depth must contribute a small, physically meaningful correction. In other words, the model is trained to progressively approach the same optimal configuration across all depths. 

Specifically, at each layer $t$ we apply the differentiable periodicity-aware Cartesian displacement loss $\mathcal{L}_{\text{pos}}$ together with a lattice loss. The overall objective is
\begin{equation}
\mathcal{L}
=
\sum_{t=1}^{T}
\alpha_t
\Bigl(
\mathcal{L}_{\text{pos}}^{(t)}
+
\sum_{c \in \mathcal{V}^l}
\bigl\|
\vec{\bm{l}}_c^{(t)} - \vec{\bm{l}}_c^{\text{gt}}
\bigr\|_1
\Bigr),
\label{eq:loss}
\end{equation}
where $\mathcal{L}_{\text{pos}}^{(t)}$ is computed using the nearest-image alignment described above, and $\alpha_t$ controls the contribution of each layer (set to $1$ by default).

\subsection{SE(3)-Equivariance Proof}
\label{sec:equivariance_proof}
Let $\Phi$ denote one interaction layer of E$^{3}$Relax-H$^{2}$. Our goal is to show that $\Phi$ is $\mathrm{SE}(3)$-equivariant, namely,
\begin{equation}
\Phi(g \cdot x)= g \cdot \Phi(x), \qquad \forall g \in \mathrm{SE}(3),
\end{equation}
where $g \cdot x$ denotes the action of a rigid motion on the input crystal and its associated features. To establish this result, we first specify the action of $\mathrm{SE}(3)$ on each type of geometric and feature variable.

\textbf{Definition 1.} \textit{Let $g=(R,\mathbf{t}) \in \mathrm{SE}(3)$, with $R\in \mathrm{SO}(3)$ and $\mathbf{t}\in\mathbb{R}^3$. The action of $g$ on atomic coordinates, lattice vectors, and node features is defined by
\begin{equation}
(\vec{\mathbf{r}}_i, \vec{\bm{l}}_c, \mathbf{h}_i, \vec{\mathbf{h}}_i, \mathbf{s}_c, \vec{\mathbf{s}}_c)
\mapsto
(R\vec{\mathbf{r}}_i + \mathbf{t}, \; R\vec{\bm{l}}_c, \; \mathbf{h}_i, \; R\vec{\mathbf{h}}_i, \; \mathbf{s}_c, \; R\vec{\mathbf{s}}_c).
\end{equation}
}

We next establish a basic technical lemma collecting the closure properties of invariant and equivariant quantities under the operations used in our architecture. 

\textbf{Lemma 1.} \textit{The following facts hold:
\begin{enumerate}[label=(\arabic*)]
    \item For any $\mathbf{x},\mathbf{y}\in\mathbb{R}^3$,
    \begin{equation}
    (R\mathbf{x})\cdot(R\mathbf{y})=\mathbf{x}\cdot\mathbf{y}
    \end{equation}
    \begin{equation}
    \|R\mathbf{x}\|=\|\mathbf{x}\|,
    \end{equation}
    \begin{equation}
    \|(R\mathbf{x})\times (R\mathbf{y})\|=\|\mathbf{x}\times \mathbf{y}\|.
    \end{equation}
    \item If $\mathbf{x}\neq \mathbf{0}$, then normalization is equivariant:
    \begin{equation}
    \frac{R\mathbf{x}}{\|R\mathbf{x}\|}
    =
    R\frac{\mathbf{x}}{\|\mathbf{x}\|}.
    \end{equation}
    We assume that whenever a normalized vector of the form $\mathbf{x}/\|\mathbf{x}\|$ is used, its denominator is nonzero. In practice, one may add a small $\varepsilon$ for numerical stability without changing the equivariance argument.
    \item If $\psi$ is an invariant scalar and $\mathbf{v}$ is an equivariant vector, then $\psi\,\mathbf{v}$ is equivariant.
    \item Sums, residual additions, and mean pooling preserve invariance/equivariance within each representation type.
    \item Any linear map acting only on feature channels, and not on the Cartesian dimension, preserves the transformation type.
\end{enumerate}
}

We first prove equivariance of the two key modules newly introduced in E$^{3}$Relax-H$^{2}$, namely H$^2$-MP and lattice-to-atom broadcast. We then briefly verify the remaining components.

\textbf{Proposition 1.}
\label{prop:h2mp_equiv}
\textit{The H$^2$-MP module maps invariant scalar inputs and equivariant vector inputs to invariant scalar outputs and equivariant vector outputs.}

\begin{proof}
For each edge $(j\to i)$, the relative displacement is $\vec{\mathbf{r}}_{ji}=\vec{\mathbf{r}}_i-\vec{\mathbf{r}}_j$.
Under $g=(R,\mathbf{t})$,
\begin{equation}
\vec{\mathbf{r}}_{ji}\mapsto
(R\vec{\mathbf{r}}_i+\mathbf{t})-(R\vec{\mathbf{r}}_j+\mathbf{t})
=
R(\vec{\mathbf{r}}_i-\vec{\mathbf{r}}_j)
=
R\vec{\mathbf{r}}_{ji}.
\end{equation}
Hence $\|\vec{\mathbf{r}}_{ji}\|$ is invariant and the unit direction
$\hat{\mathbf{r}}_{ji}$ is equivariant by Lemma~1 (1)-(2).

The local reference direction is
\begin{equation}
\hat{\mathbf{r}}_{\mathrm{ref},i}
=
\frac{\sum_{k\in\mathcal{N}(i)} \hat{\mathbf{r}}_{ki}}
{\left\|\sum_{k\in\mathcal{N}(i)} \hat{\mathbf{r}}_{ki}\right\|}.
\end{equation}
Since each $\hat{\mathbf{r}}_{ki}$ transforms as $R\hat{\mathbf{r}}_{ki}$, their sum transforms by left multiplication with $R$, and therefore $\hat{\mathbf{r}}_{\mathrm{ref},i}$ is equivariant.

For each degree $\ell$, define
\begin{equation}
\mathbf{Y}^{\ell}(\hat{\mathbf{r}})
=
\bigl(Y_{\ell m}(\hat{\mathbf{r}})\bigr)_{m=-\ell}^{\ell}
\end{equation}
Using the Wigner-$D$ transformation law,
\begin{equation}
\mathbf{Y}^{\ell}(R\hat{\mathbf{r}})
=
D^{\ell}(R)\,\mathbf{Y}^{\ell}(\hat{\mathbf{r}}).
\end{equation}
Since $D^\ell(R)$ is orthogonal in the real basis, the bilinear quantity
\begin{equation}
z_{ji}^{\ell}
=
\sum_{m=-\ell}^{\ell}
Y_{\ell m}(\hat{\mathbf{r}}_{ji})
Y_{\ell m}(\hat{\mathbf{r}}_{\mathrm{ref},i})
=
\bigl(\mathbf{Y}^{\ell}(\hat{\mathbf{r}}_{ji})\bigr)^\top
\mathbf{Y}^{\ell}(\hat{\mathbf{r}}_{\mathrm{ref},i})
\end{equation}
is rotation-invariant. It is also translation-invariant because translations cancel in $\vec{\mathbf{r}}_{ji}$. Therefore the concatenated descriptor $\mathbf{z}_{ji} = \bigoplus_{\ell=0}^{\ell_{\max}} z_{ji}^{\ell}$ is $\mathrm{SE}(3)$-invariant.

Since $\mathbf{z}_{ji}$ and $\|\vec{\mathbf{r}}_{ji}\|$ are invariant, the quantities $\mathbf{w}_{ji}$, $\mathbf{e}_{ji}$, and $\phi_{\mathbf{h}}(\mathbf{h}_j)$ are all invariant. Therefore $\mathbf{m}_{ji}$ and its aggregation $\mathbf{m}_i$ are invariant. Likewise, $\phi_{\vec{\mathbf{h}}}(\mathbf{m}_{ji})$ and $\phi_{\vec{\mathbf{r}}}(\mathbf{m}_{ji})$ are invariant scalar gates, while $\vec{\mathbf{h}}_j$ and $\hat{\mathbf{r}}_{ji}$ are equivariant vectors, so each term in $\vec{\mathbf{m}}_i$ is equivariant. Hence $\vec{\mathbf{m}}_i$ is equivariant.
\end{proof}

\textbf{Proposition 2.} 
\label{prop:l2a_equiv}
\textit{The lattice-to-atom broadcast stage in LA-MP is $\mathrm{SE}(3)$-equivariant.}
\begin{proof}
We first show that the centered coordinates are invariant under translations and equivariant under rotations. Define
\begin{equation}
\bar{\vec{\mathbf{r}}}
=
\frac{1}{N}\sum_{i=1}^{N}\vec{\mathbf{r}}_i,
\qquad
\tilde{\vec{\mathbf{r}}}_i
=
\vec{\mathbf{r}}_i-\bar{\vec{\mathbf{r}}}.
\end{equation}
Under $g=(R,\mathbf{t})$,
\begin{align}
\bar{\vec{\mathbf{r}}}
&\mapsto
\frac{1}{N}\sum_{i=1}^N (R\vec{\mathbf{r}}_i+\mathbf{t})
=
R\bar{\vec{\mathbf{r}}}+\mathbf{t}, \\
\tilde{\vec{\mathbf{r}}}_i
&\mapsto
(R\vec{\mathbf{r}}_i+\mathbf{t})-(R\bar{\vec{\mathbf{r}}}+\mathbf{t})
=
R(\vec{\mathbf{r}}_i-\bar{\vec{\mathbf{r}}})
=
R\tilde{\vec{\mathbf{r}}}_i.
\end{align}
Thus $\tilde{\vec{\mathbf{r}}}_i$ is invariant under translations and equivariant under rotations.

Now consider the geometric descriptors $d_{ic}^{\parallel}$ and $d_{ic}^{\perp}$. Under rotation,
\begin{align}
d_{ic}^{\parallel}
&\mapsto
\frac{(R\tilde{\vec{\mathbf{r}}}_i)\cdot(R\vec{\bm{l}}_c)}{\|R\vec{\bm{l}}_c\|}
=
\frac{\tilde{\vec{\mathbf{r}}}_i\cdot\vec{\bm{l}}_c}{\|\vec{\bm{l}}_c\|}
=
d_{ic}^{\parallel}, \\
d_{ic}^{\perp}
&\mapsto
\frac{\|(R\tilde{\vec{\mathbf{r}}}_i)\times(R\vec{\bm{l}}_c)\|}{\|R\vec{\bm{l}}_c\|}
=
\frac{\|\tilde{\vec{\mathbf{r}}}_i\times\vec{\bm{l}}_c\|}{\|\vec{\bm{l}}_c\|}
=
d_{ic}^{\perp},
\end{align}
using Lemma 1 (1). Since $\tilde{\vec{\mathbf{r}}}_i$ is translation-invariant, both descriptors are also translation-invariant. Therefore the lattice-atom encoding $\mathbf{e}_{ci}
=
\phi_{\mathrm{enc}}
\bigl(\lambda(d_{ic}^{\parallel})\oplus \lambda(d_{ic}^{\perp})\bigr)$ is invariant.

Next, recall that $\hat{\mathbf{r}}_{ci}
= {(\vec{\bm{l}}_c - \tilde{\vec{\mathbf{r}}}_i)} /
{\|\vec{\bm{l}}_c - \tilde{\vec{\mathbf{r}}}_i\|}$. Because both $\vec{\bm{l}}_c$ and $\tilde{\vec{\mathbf{r}}}_i$ transform by left multiplication with $R$,
\begin{equation}
\vec{\bm{l}}_c-\tilde{\vec{\mathbf{r}}}_i
\mapsto
R(\vec{\bm{l}}_c-\tilde{\vec{\mathbf{r}}}_i),
\end{equation}
and thus $\hat{\mathbf{r}}_{ci}$ is equivariant.

In Eqn.~\eqref{eq:la_s}, all inputs $\mathbf{h}_i^{(t)}$, $\mathbf{s}_c^{(t)}$, and $\mathbf{e}_{ci}^{(t)}$ are invariant, and $\phi_s^a,\phi_s^l$ are invariant MLPs. Hence each summand is invariant, and so are the sum and residual update. Therefore $\mathbf{h}_i^{+}$ is invariant.

In Eqn.~\eqref{eq:la_v}, $\mathbf{h}_i^{+}$ is invariant, so $\phi_v^a(\mathbf{h}_i^{+})$ and $\phi_v^r(\mathbf{h}_i^{+})$ are invariant scalar gates. Thus the gated terms
$\phi_{v}^{a}(\mathbf{h}_i^{+})\circ \vec{\mathbf{h}}_i^{(t)}$ and
$\phi_{v}^{r}(\mathbf{h}_i^{+})\circ \hat{\mathbf{r}}_{ci}^{(t)}$
are equivariant. The term $\mathbf{W}_c\,\vec{\mathbf{s}}_c^{(t)}$ is also equivariant because $\mathbf{W}_c$ acts only on feature channels. Therefore the full update, including the residual connection, remains equivariant, and hence $\vec{\mathbf{h}}_i^{+}$ is equivariant.
\end{proof}

The remaining modules are also $\mathrm{SE}(3)$-equivariant. In self-interaction, $\|\mathbf{U}\,\vec{\mathbf{m}}_i\|$ is rotation-invariant, where the norm is taken over the Cartesian dimension, so the scalar update is invariant. The vector update remains equivariant because $\mathbf{V}$ acts only on feature channels. In atom-to-lattice aggregation, mean pooling preserves transformation type, and the channel-wise map $\phi_{a\rightarrow l}^{v}$ preserves equivariance. In lattice-lattice interaction, the matrices $\mathbf{W}_s,\mathbf{W}_v$ mix only the lattice-axis index and therefore preserve scalar/vector type. Finally, in structure updating, the bias-free linear projections $\mathbf{W}_p,\mathbf{W}_l$ map equivariant vector features to equivariant geometric displacements, so the residual updates of atomic coordinates and lattice vectors are $\mathrm{SE}(3)$-equivariant. Hence one full interaction layer is $\mathrm{SE}(3)$-equivariant, and by composition so is the full network.

\subsection{Performance Indicators}
\label{sec:metrics}
We evaluate structural accuracy using three metrics: mean absolute error (MAE) of atomic coordinates, cell shape deviation, and MAE of cell volume. The coordinate MAE measures atomic positional errors, while the latter two metrics quantify deviations in lattice geometry.

\subsubsection{MAE of Coordinates}
The coordinate MAE measures the average Cartesian displacement between the predicted and DFT-relaxed atomic positions:
\begin{equation}
\Delta_{\rm coord}
=
\frac{1}{3N}
\sum_{i \in \mathcal{V}^a}
\left\lvert
\vec{\mathbf{r}}_i^{\mathrm{align}} - \vec{\mathbf{r}}_i^{\mathrm{gt}}
\right\rvert.
\end{equation}
where $N$ denotes the number of atoms, $\vec{\mathbf{r}}_i^{\mathrm{gt}}$ is the ground-truth Cartesian coordinate of atom $i$, and $\vec{\mathbf{r}}_i^{\mathrm{align}}$ is the predicted coordinate after nearest-image alignment under periodic boundary conditions.

\subsubsection{Cell Shape Deviation}
Cell shape deviation is quantified using the Frobenius norm of the difference between the predicted and reference lattice metric tensors. Given the lattice matrix $\mathbf{L}$, the metric tensor is defined as $\mathbf{G} = \mathbf{L}^\top \mathbf{L}$. The deviation is computed as
\begin{equation}
\Delta_{\rm shape}
=
\left\|
\mathbf{G}^{\mathrm{pred}} - \mathbf{G}^{\mathrm{gt}}
\right\|_F,
\end{equation}
where $\mathbf{G}^{\mathrm{pred}}$ and $\mathbf{G}^{\mathrm{gt}}$ denote the predicted and DFT-relaxed lattice metric tensors.

\subsubsection{MAE of Cell Volume}
The cell volume MAE evaluates the accuracy of the predicted unit-cell scale and is defined as the absolute difference between the predicted and reference volumes:
\begin{equation}
\Delta_{\rm volume}
=
\left|
\det(\mathbf{L}^{\mathrm{pred}}) - \det(\mathbf{L}^{\mathrm{gt}})
\right|,
\end{equation}
where $\mathbf{L}^{\mathrm{pred}}$ and $\mathbf{L}^{\mathrm{gt}}$ denote the predicted and DFT-relaxed lattice matrices, respectively.

\subsubsection{Match Rate}
We utilize the \texttt{StructureMatcher} function from the Pymatgen package~\cite{ong2013python} to compare the predicted structure with the corresponding DFT-relaxed ground-truth structure. The matcher evaluates structural equivalence by considering lattice similarity, atomic species correspondence, and fractional coordinate alignment. A structure is considered successfully matched if the predicted structure can be mapped onto the DFT-relaxed reference within the specified tolerances. Default parameters are adopted (ltol = 0.2, stol = 0.3) to ensure consistent and objective comparisons across all datasets. The match rate is then computed as the fraction of successfully matched structures over the total number of evaluated samples.

\subsection{Implementation Details}
E$^{3}$Relax-H$^{2}$ is implemented in PyTorch (with e3nn for spherical harmonics) and trained on a single NVIDIA L40S GPU with 48 GB of memory. We use a hidden feature width $F = 512$ and stack $T = 4$ interaction layers.

We build periodic neighbor graphs using a distance cutoff $D = 6.0\,\mathrm{\AA}$ with a maximum number of neighbors $K = 50$. Interatomic distances are expanded using a radial basis with 128 channels and a polynomial cutoff envelope with exponent 5. We use a Gaussian radial basis.

Within the H$^2$-MP module, spherical harmonics are evaluated up to degree $\ell_{\max}=3$. Degree-wise bilinear invariants are obtained by contracting the spherical harmonic coefficients of each edge direction with those of the corresponding local reference direction, yielding $(\ell_{\max}+1)$ scalar invariants per edge. These invariants are then normalized using LayerNorm. The resulting geometric descriptors are passed through an MLP to produce $\tanh$-bounded gates that modulate the radial edge features.

Model parameters are optimized using AdamW with an initial learning rate of $1\times10^{-4}$. A plateau-based learning rate scheduler is employed, reducing the learning rate when the monitored validation metric fails to improve for 5 consecutive epochs.

\begin{table*}[t]
\centering
\caption{Summary of the benchmark datasets used in our experiments.}
\begin{tabular}{lllll}
\hline
Dataset & Dimensionality & Sample size & Elements covered & Split ratio (Train : Val : Test) \\ \hline
Materials Project (MP) \cite{jain2013commentary, chen2022universal}
& 3D & 62\,724 & 89 & 8 : 1 : 1 \\
X-Mn-O \cite{kim2023structure}
& 3D & 28\,579 & 6 & 8 : 1 : 1 \\
JARVIS DFT\cite{choudhary2020joint} & 3D & 33280 & 89 & 8 : 1 : 1 \\
OC20 \cite{chanussot2021open}
& 3D & 560\,182 & 55 & 0.82 : 0.045 : 0.045 : 0.045 : 0.045$^{*}$ \\
C2DB \cite{gjerding2021recent, lyngby2022data}
& 2D & 11\,581 & 62 & 6 : 2 : 2 \\
vdW heterostructures
& 2D & 3706 & 61 & 8 : 1 : 1 \\ \hline
\end{tabular}
\label{tbl:datasets}
\vspace{0.5em}

\footnotesize
\raggedright
$^{*}$Train: Val\_ID: Val\_OOD\_Ads: Val\_OOD\_Cat: Val\_OOD\_Both, where Val\_ID denotes in-distribution validation samples, and Val\_OOD\_Ads, Val\_OOD\_Cat, and Val\_OOD\_Both correspond to out-of-distribution validation splits involving adsorbates, catalysts, and their combination, respectively.
\end{table*}

\begin{figure}[t]
  \centering
  \includegraphics[width=1.0\columnwidth]{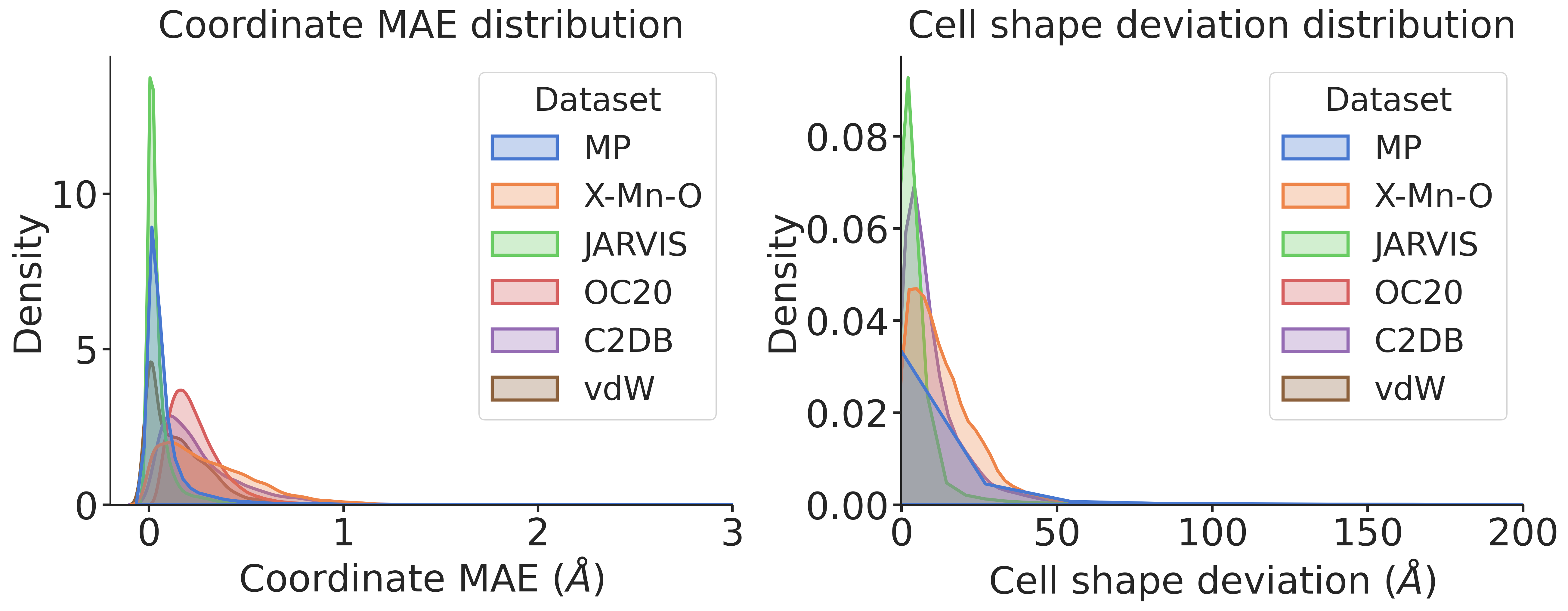}
  \caption{Distributions of structural deviations between unrelaxed and DFT-relaxed structures across the six benchmark datasets. The left panel shows atomic coordinate MAE, and the right panel shows cell-shape deviation.}
  \label{fgr:dataset_distribution}
\end{figure}

\section{Experiments}
\subsection{Datasets}
We evaluate E$^{3}$Relax-H$^{2}$ on six benchmark datasets: Materials Project (MP) \cite{jain2013commentary, chen2022universal}, X-Mn-O (X = Mg, Ca, Ba, Sr) \cite{kim2023structure}, JARVIS DFT \cite{choudhary2020joint}, OC20 \cite{chanussot2021open}, C2DB \cite{gjerding2021recent, lyngby2022data}, and a curated set of layered van der Waals (vdW) heterostructures. The vdW heterostructures are derived from C2DB by stacking relaxed 2D monolayers into bilayer van der Waals heterostructures, which are subsequently DFT-relaxed to obtain reference geometries. Each data point consists of a paired structure, an unrelaxed initial structure and its corresponding DFT-relaxed geometry. Dataset statistics are summarized in Table~\ref{tbl:datasets}.

Figure~\ref{fgr:dataset_distribution} summarizes the distributions of structural deviations between unrelaxed and DFT-relaxed structures across the six benchmarks. In terms of coordinate MAE, MP, JARVIS and vdW exhibit small deviations, whereas X-Mn-O and OC20 show broader distributions with heavier tails. Note that for OC20 and vdW, the lattice parameters do not change before and after structural optimization. 

Overall, these datasets include both 2D and 3D materials, covering a wide range of chemical compositions, bonding environments, and structural complexities. They include cases with minor and significant distortions, providing a comprehensive and challenging benchmark for evaluating structure relaxation models.

\subsection{Baselines}
We compare E$^{3}$Relax-H$^{2}$  with SOTA iteration-free models, including E$^{3}$Relax \cite{yang2026equivariant}, DeepRelax \cite{yang2024scalable}, EquiformerV2 \cite{equiformer_v2}, HEGNN \cite{cen2024high}, GotenNet \cite{aykent2025gotennet}, PAINN \cite{schutt2021equivariant}, and EGNN \cite{satorras2021n}. All baseline models are implemented using the official source code released by the original authors and are evaluated under the same training, validation, and testing splits for fair comparison. We repeat each experiment three times and report average results for all datasets except OC20, which is too large and requires substantial training time. The Dummy model, which simply returns the input structure, serves as a control. A brief introduction to the baseline models is as follows.
\begin{itemize}
    \item \textbf{E$^{3}$Relax} is our prior work published at AAAI-26~\cite{yang2026equivariant}, which we extend in this paper to E$^{3}$Relax-H$^{2}$.
    \item \textbf{DeepRelax} predicts relaxed pairwise distances and lattice parameters, then reconstructs Cartesian coordinates via Euclidean distance geometry.
    \item \textbf{PAINN} and \textbf{EGNN} are lightweight equivariant GNNs. We adapt their force heads to output atomic displacements and follow DeepRelax for lattice prediction.
    \item \textbf{EquiformerV2} is an equivariant Transformer. We adapt its force head to output atomic displacements and use the DeepRelax lattice-prediction pipeline.
    \item \textbf{HEGNN} extends EGNN with high-degree steerable representations via efficient scalarization.
    \item \textbf{GotenNet} combines geometric tensor attention with hierarchical refinement for efficient 3D equivariant modeling.
\end{itemize}

\begin{table*}[!ht]
\centering
\caption{Comparison of E$^{3}$Relax-H$^{2}$ with baseline models on two 3D materials datasets (MP and X-Mn-O). Metrics are MAE of atomic coordinates ($\mathrm{\AA}$), cell shape deviation ($\mathrm{\AA}^2$), MAE of cell volume ($\mathrm{\AA^{3}}$), and structure match rate between ML-relaxed and DFT-relaxed structures. The best results are in bold, the second-best are underlined. The improvement (\%) row reports the percentage gain of E$^{3}$Relax-H$^{2}$ over the second-best model.}
\setlength{\tabcolsep}{3.5pt}
\begin{tabular}{lcccccccc}
\toprule
\multirow{2}{*}{Model} &
\multicolumn{4}{c}{MP} &
\multicolumn{4}{c}{X-Mn-O} \\ 
\cmidrule(lr){2-5}\cmidrule(lr){6-9}
& MAE Coords.$\downarrow$ & Shape Dev.$\downarrow$ & MAE Vol.$\downarrow$ & Struct. Match$\uparrow$
& MAE Coords.$\downarrow$ & Shape Dev.$\downarrow$ & MAE Vol.$\downarrow$ & Struct. Match$\uparrow$ \\ 
\midrule

Dummy        & 0.095 & 11.502 & 27.006 & 0.957 
             & 0.314 & 13.913 & 32.839 & 0.648 \\

PAINN        & 0.088 & 4.775 & 9.343 & 0.942
             & 0.159 & 3.784 & 3.803 & 0.757 \\

EGNN         & 0.086 & 4.834 & 9.253 & 0.941
             & 0.166 & 3.814 & 4.208 & 0.738 \\

HEGNN        & 0.091 & 4.987 & 10.382 & 0.946
             & 0.191 & 3.862 & 4.421 & 0.754 \\

GotenNet     & 0.078 & 4.111 & 7.456 & 0.940
             & 0.160 & 3.669 & 3.559 & 0.754 \\

EquiformerV2 & 0.069 & 4.329 & 8.539 & 0.945
             & 0.115 & 3.558 & 3.713 & 0.828 \\

DeepRelax    & 0.066 & 4.554 & 9.611 & 0.944
             & 0.116 & 3.530 & 3.442 & 0.847 \\

E$^{3}$Relax 
& \underline{0.057} & \underline{4.020} & \underline{7.417} & \underline{0.960}
& \underline{0.105} & \underline{3.447} & \underline{3.369} & \underline{0.858} \\

E$^{3}$Relax-H$^{2}$ 
& \textbf{0.054} & \textbf{3.904} & \textbf{7.307} & \textbf{0.962}
& \textbf{0.092} & \textbf{3.265} & \textbf{3.191} & \textbf{0.872} \\

\midrule
Improvement 
& 5.26\% & 2.89\% & 1.48\% & 0.21\%
& 12.38\% & 5.28\% & 5.28\% & 1.63\% \\
\bottomrule
\end{tabular}
\label{tbl:mp_xmno}
\end{table*}

\begin{table}[!ht]
\centering
\caption{Comparison of E$^{3}$Relax-H$^{2}$ with baseline models on the JARVIS dataset. }
\setlength{\tabcolsep}{2pt}
\begin{tabular}{lcccc}
\toprule
Model & MAE Coords.$\downarrow$ & Shape Dev.$\downarrow$ & MAE Vol.$\downarrow$ & Struct. Match$\uparrow$ \\
\midrule

Dummy 
& 0.058 & 6.291 & 12.307 & 0.959 \\

PAINN 
& 0.050 & 4.161 & 6.482 & 0.957 \\

EGNN 
& 0.055 & 4.489 & 6.201 & 0.955 \\

HEGNN 
& 0.055 & 4.212 & 6.023 & 0.956 \\

GotenNet 
& 0.050 & 3.990 & 5.650 & 0.955 \\

EquiformerV2 
& 0.047 & \underline{3.837} & 5.394 & 0.955 \\

DeepRelax 
& 0.050 & 3.884 & 5.226 & \underline{0.960} \\

E$^{3}$Relax 
& \underline{0.043} & 3.865 & 6.201 & \underline{0.960} \\

E$^{3}$Relax-H$^{2}$ 
& \textbf{0.040} & \textbf{3.665} & \textbf{5.003} & \textbf{0.961} \\

\bottomrule
\end{tabular}
\label{tbl:jarvis}
\end{table}

\begin{table*}[!ht]
\centering
\caption{Performance comparison on the OC20 dataset. Coordinate MAE is measured on movable atoms. Val ID denotes in-distribution validation samples. Val OOD Ads, Val OOD Cat, and Val OOD Both correspond to out-of-distribution splits involving adsorbates, catalysts, and their combination. Best results are in bold and second-best are underlined. Training and inference time are measured on a single NVIDIA L40S GPU.}
\begin{tabular}{lcccccccccc}
\toprule
\multirow{2}{*}{Model} &
\multicolumn{2}{c}{Val ID} &
\multicolumn{2}{c}{Val OOD Ads} &
\multicolumn{2}{c}{Val OOD Cat} &
\multicolumn{2}{c}{Val OOD Both} &
\multirow{2}{*}{Train (h)} &
\multirow{2}{*}{Infer (ms)} \\
\cmidrule(lr){2-3}
\cmidrule(lr){4-5}
\cmidrule(lr){6-7}
\cmidrule(lr){8-9}
& MAE$\downarrow$ & Match$\uparrow$
& MAE$\downarrow$ & Match$\uparrow$
& MAE$\downarrow$ & Match$\uparrow$
& MAE$\downarrow$ & Match$\uparrow$
& & \\
\midrule

Dummy        
& 0.245 & 0.244 
& 0.240 & 0.284 
& 0.247 & 0.229 
& 0.201 & 0.322 
& - & - \\

PAINN        
& 0.161 & 0.490 
& 0.166 & 0.467 
& 0.164 & 0.484 
& 0.137 & 0.546 
& 9.31 & 1.80 \\

EGNN         
& 0.174 & 0.463 
& 0.178 & 0.436 
& 0.175 & 0.459 
& 0.148 & 0.514 
& 8.74 & 1.83 \\

HEGNN        
& 0.178 & 0.441 
& 0.184 & 0.408 
& 0.177 & 0.436 
& 0.151 & 0.482 
& 32.60 & 3.19 \\

GotenNet     
& 0.156 & 0.515 
& 0.162 & 0.479 
& 0.157 & 0.507 
& 0.132 & 0.564 
& 21.05 & 5.11 \\

EquiformerV2 
& \textbf{0.136} & \textbf{0.580} 
& \textbf{0.141} & \textbf{0.552} 
& \textbf{0.138} & \textbf{0.574} 
& \textbf{0.113} & \textbf{0.641} 
& 91.70 & 38.62 \\

DeepRelax    
& 0.172 & 0.451 
& 0.175 & 0.431 
& 0.176 & 0.443 
& 0.146 & 0.504 
& 34.89 & 89.44 \\

E$^{3}$Relax 
& 0.156 & 0.500
& 0.160 & 0.475
& 0.156 & 0.498
& 0.129 & 0.564
& 9.10 & 8.12 \\

E$^{3}$Relax-H$^{2}$ 
& \underline{0.149} & \underline{0.523}
& \underline{0.155} & \underline{0.491}
& \underline{0.151} & \underline{0.519}
& \underline{0.126} & \underline{0.577}
& 10.56 & 10.11 \\

\bottomrule
\end{tabular}
\label{tbl:oc20}
\end{table*}

\begin{figure}[t]
  \centering
  \includegraphics[width=1.0\columnwidth]{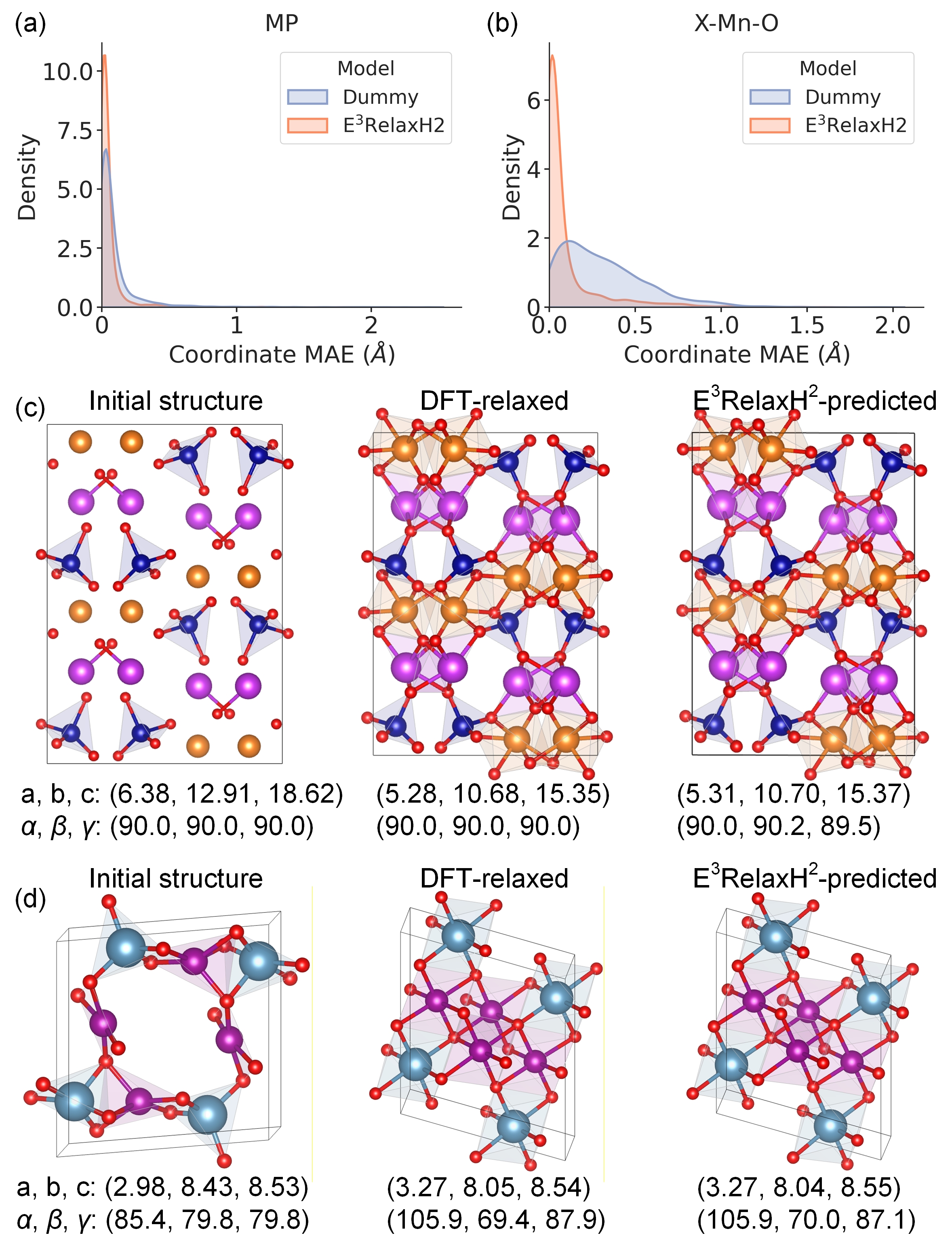}
  \caption{Distribution of coordinate MAE ($\mathrm{\AA}$) between predicted and DFT-relaxed structures for Dummy and E$^{3}$Relax-H$^{2}$ on MP and X-Mn-O. (a) MP. (b) X-Mn-O. (c) Example prediction on MP ($\mathrm{Bi_8Cr_8Mg_8O_{40}}$). (d) Example prediction on X-Mn-O ($\mathrm{Ca_4Mn_4O_8}$). Lattice constants \(a\), \(b\), and \(c\) are reported in angstroms (Å), and lattice angles \(\alpha\), \(\beta\), and \(\gamma\) in degrees ($^\circ$).}
  \label{fgr:performance_mp_xmno}
\end{figure}

\subsection{Results on 3D Materials Datasets}
We first evaluate E$^{3}$Relax-H$^{2}$ on four 3D benchmarks: MP, X-Mn-O, JARVIS, and OC20. Table~\ref{tbl:mp_xmno} summarizes the results on MP and X-Mn-O, where E$^{3}$Relax-H$^{2}$ achieves the best performance across all four metrics, confirming its ability to jointly model atomic displacements and lattice deformations.

Figure~\ref{fgr:performance_mp_xmno} (a) and (b) compares the coordinate-MAE distributions of the Dummy baseline and E$^{3}$Relax-H$^{2}$ on MP and X-Mn-O. In both datasets, E$^{3}$Relax-H$^{2}$ shifts the distributions toward smaller errors, indicating more consistent predictions that are closer to the DFT-relaxed geometries. Notably, the initial structures in MP and X-Mn-O exhibit different levels of structural deviation, as reflected by the MAE distributions of the Dummy model in Figure~\ref{fgr:performance_mp_xmno} (a) and (b). Despite these notably different data distributions, E$^{3}$Relax-H$^{2}$ consistently shifts the error distributions toward lower values in both cases, demonstrating stable performance across datasets with varying degrees of initial structural distortion. Figure~\ref{fgr:performance_mp_xmno} (c) and (d) presents two representative structures predicted by E$^{3}$Relax-H$^{2}$. The predicted geometries closely match the DFT-relaxed references, demonstrating that the model captures both atomic displacements and lattice deformations well.

Table~\ref{tbl:jarvis} reports results on the JARVIS dataset. Compared with E$^{3}$Relax, E$^{3}$Relax-H$^{2}$ consistently improves structural accuracy across all metrics. Specifically, E$^{3}$Relax-H$^{2}$ achieves a relative reduction of approximately 7.0\% in coordinate MAE, 5.17\% in shape deviation, and 19.32\% in cell volume MAE compared with E$^{3}$Relax, indicating consistent refinement of both atomic positions and lattice geometry.

Table~\ref{tbl:oc20} reports results on the large-scale OC20 dataset. E$^{3}$Relax-H$^{2}$ consistently outperforms PAINN, EGNN, HEGNN, GotenNet, DeepRelax, and E$^{3}$Relax across all validation splits. Although EquiformerV2 achieves the lowest coordinate MAE, it incurs substantially higher training and inference costs. In contrast, E$^{3}$Relax-H$^{2}$ offers a more favorable accuracy--efficiency trade-off. Because lattice parameters remain unchanged in OC20, this benchmark does not fully reflect the advantage of E$^{3}$Relax-H$^{2}$'s explicit lattice modeling, which is designed to handle joint atomic-lattice relaxation.

\begin{figure}[tb]
  \centering
  \includegraphics[width=1.0\columnwidth]{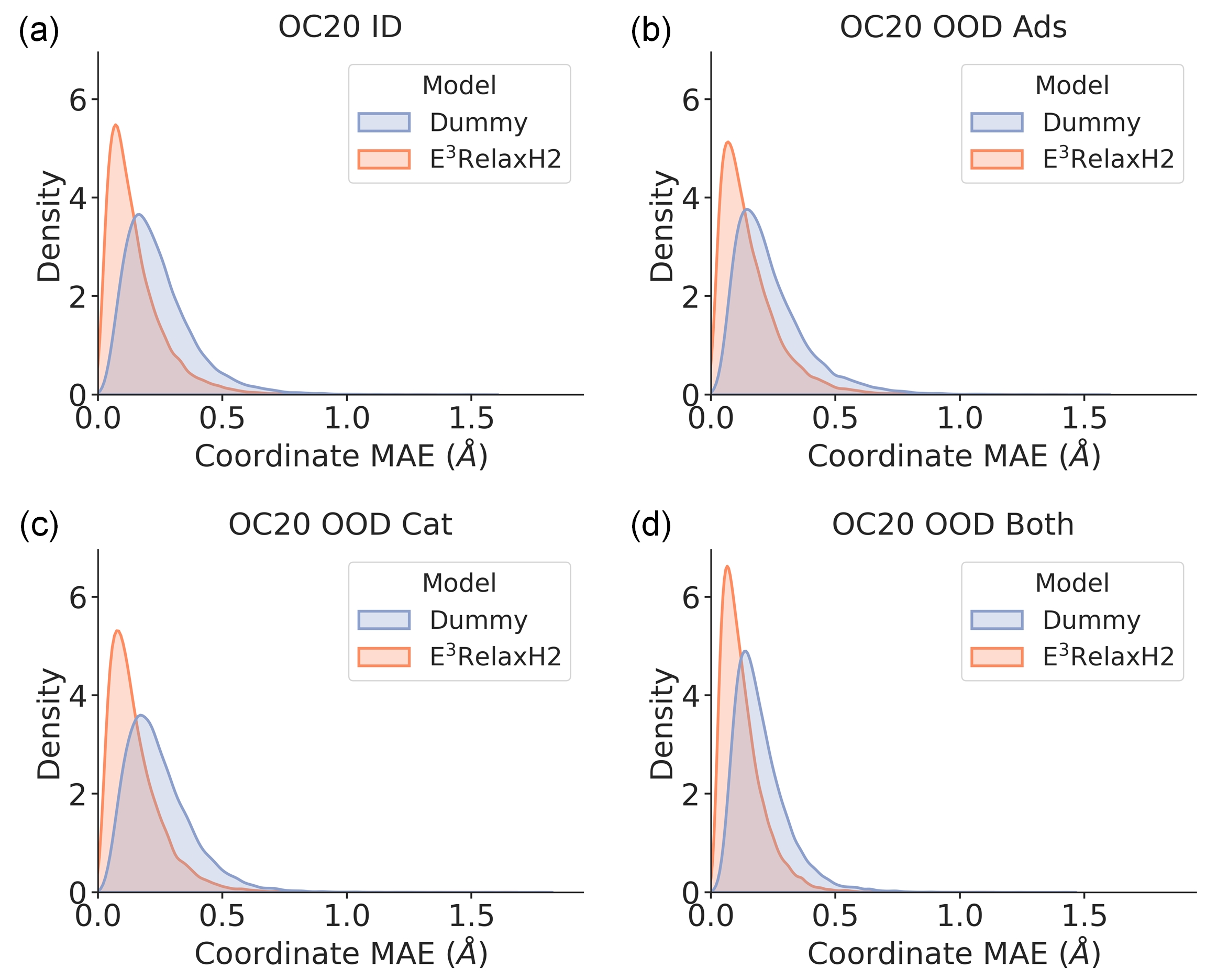}
  \caption{Distribution of coordinate MAE ($\mathrm{\AA}$) between predicted and DFT-relaxed structures for Dummy and E$^{3}$Relax-H$^{2}$ on OC20. (a) ID. (b) OOD Ads. (c) OOD Cat. (d) OOD Both.}
  \label{fgr:performance_oc20}
\end{figure}

Interestingly, all iteration-free models, including E$^{3}$Relax-H$^{2}$, exhibit similar coordinate-MAE distributions across ID and OOD splits (Table~\ref{tbl:oc20}, Figure~\ref{fgr:performance_oc20}), suggesting limited performance degradation under compositional shift. We attribute this robustness to the fact that the OC20 OOD splits primarily reflect changes in adsorbates and catalysts, rather than systematically more challenging geometric distortions.

\begin{table*}[!ht]
\centering
\caption{Comparison of E$^{3}$Relax-H$^{2}$ with baseline models on the two 2D materials datasets (C2DB and vdW heterostructures). }
\begin{tabular}{lcccccc}
\toprule
\multirow{2}{*}{Model} &
\multicolumn{4}{c}{C2DB} &
\multicolumn{2}{c}{vdW heterostructures} \\ 
\cmidrule(lr){2-5}\cmidrule(lr){6-7}
& MAE Coords.$\downarrow$ & Shape Dev.$\downarrow$ & MAE Vol.$\downarrow$ & Struct. Match$\uparrow$
& MAE Coords.$\downarrow$ & Struct. Match$\uparrow$ \\ 
\midrule

Dummy        & 0.268 & 10.615 & 149.648 & 0.723 & 0.149 & 0.901 \\
PAINN        & 0.226 & 5.952 & 61.897 & 0.792 & 0.134 & 0.913 \\
EGNN         & 0.232 & 6.065 & 67.892 & 0.770 & 0.147 & 0.905 \\
HEGNN        & 0.244 & 6.329 & 71.625 & 0.787 & 0.148 & 0.908 \\
GotenNet     & 0.208 & 5.704 & \underline{55.800} & 0.802 & 0.129 & \textbf{0.923} \\
EquiformerV2 & 0.188 & 5.660 & 60.435 & 0.811 & \textbf{0.114} & 0.918 \\
DeepRelax    & 0.196 & 5.752 & 60.216 & 0.792 & 0.166 & 0.892 \\

E$^{3}$Relax 
& \underline{0.171} & \underline{5.430} & 56.909 & \underline{0.823}
& \underline{0.124} & 0.913 \\

E$^{3}$Relax-H$^{2}$ 
& \textbf{0.161} & \textbf{5.139} & \textbf{52.772} & \textbf{0.845}
& \textbf{0.114} & \underline{0.922} \\

\bottomrule
\end{tabular}
\label{tbl:2D_performance}
\end{table*}

\subsection{Results on 2D Materials Datasets}

We further evaluate the model on 2D materials using the C2DB and vdW heterostructures datasets. As reported in Table~\ref{tbl:2D_performance}, E$^{3}$Relax-H$^{2}$ achieves the best performance on C2DB across all three metrics. In particular, it reduces coordinate MAE, cell-shape deviation, and volume MAE by 5.85\%, 5.36\%, and 5.42\%, respectively, relative to the second-best baseline. These improvements indicate that explicit joint modeling of atomic displacements and lattice deformation remains beneficial in 2D systems.

On the vdW heterostructures dataset, we make two observations. First, due to the relatively small sample size, all models achieve only modest improvements over the Dummy baseline. Second, E$^{3}$Relax-H$^{2}$ and EquiformerV2 achieve the best overall performance, with E$^{3}$Relax-H$^{2}$ obtaining the highest structure matching rate while maintaining competitive coordinate accuracy.

Overall, the results on both 2D datasets confirm that E$^{3}$Relax-H$^{2}$ generalizes well across dimensionalities, delivering consistent gains in structurally diverse settings while remaining competitive even in scenarios where lattice parameters are fixed.

\begin{figure*}[ht]
  \centering
  \includegraphics[width=1.8\columnwidth]{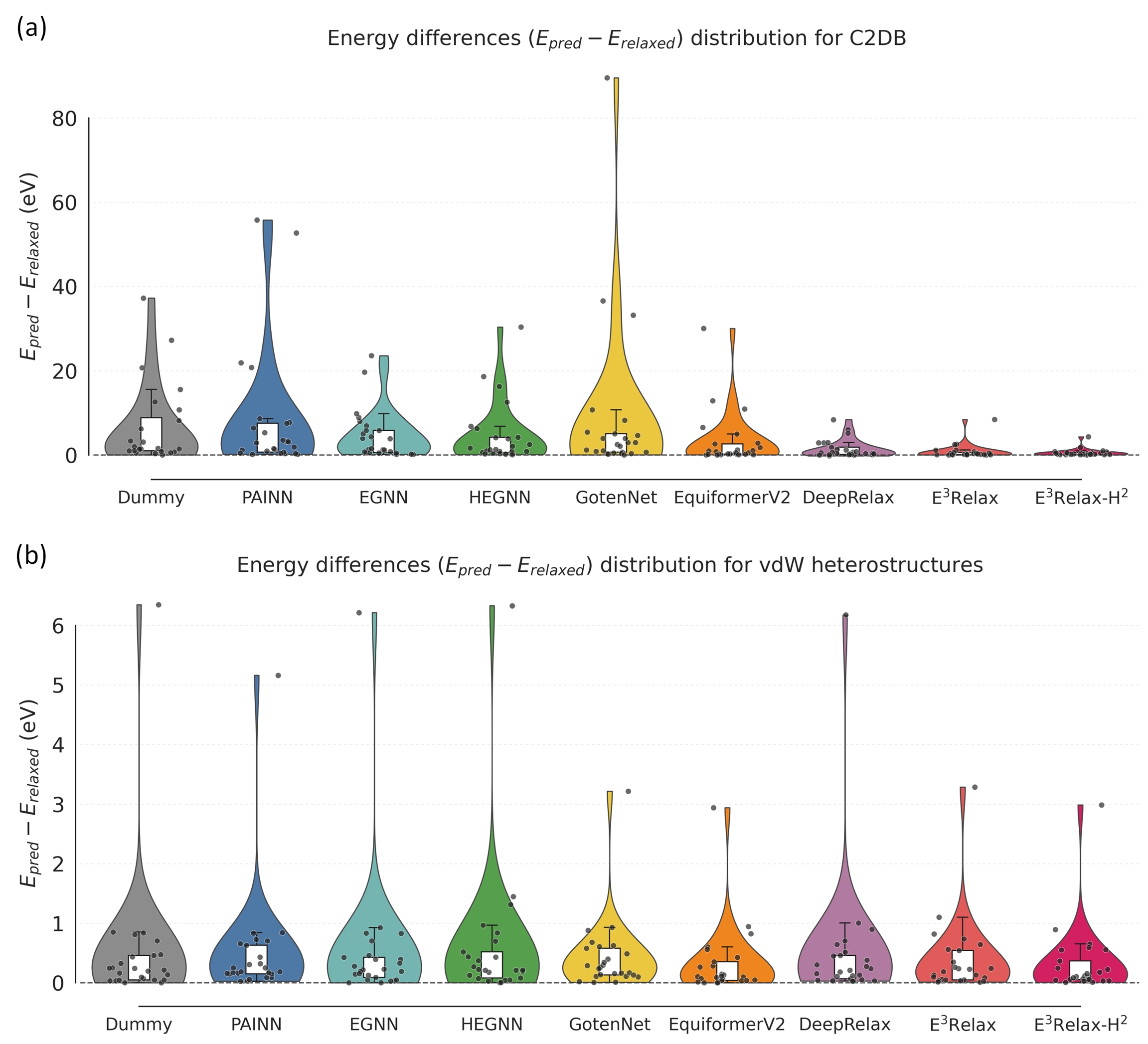}
  \caption{DFT validation on 2D materials. Distribution of total energy differences $E_{\mathrm{pred}} - E_{\mathrm{relaxed}}$ (in eV) between ML-predicted structures and DFT-relaxed references for 25 randomly selected test structures. (a) C2DB dataset. (b) vdW heterostructures dataset.}
  \label{fgr:DFT_val}
\end{figure*}

\subsection{DFT Validation}
\label{sec:dft_val}
To further validate the structural accuracy of E$^{3}$Relax-H$^{2}$, we perform post hoc DFT calculations on ML-predicted structures and compare their total energies with those of the corresponding DFT-relaxed references. For each dataset, we evaluate three types of structures: (i) the initial unrelaxed configurations, (ii) the ground-truth DFT-relaxed structures, and (iii) the ML-predicted relaxed structures produced by all baseline models.

Due to the high computational cost of DFT calculations, we randomly select 25 structures from the test sets of the two 2D materials datasets (C2DB and vdW heterostructures). These datasets are moderate in size, allowing full DFT validation while keeping the overall computational expense manageable.

Figures~\ref{fgr:DFT_val}(a) and (b) present the distributions of energy differences $E_{\mathrm{pred}} - E_{\mathrm{relaxed}}$ for C2DB and vdW heterostructures, respectively. On C2DB (Figure~\ref{fgr:DFT_val}(a)), E$^{3}$Relax-H$^{2}$ achieves the lowest median energy error and the narrowest error distribution among all methods. In particular, it significantly suppresses the heavy-tailed high-energy outliers observed in several baselines, demonstrating improved stability and robustness in predicting low-energy configurations.

On the vdW heterostructures dataset (Figure~\ref{fgr:DFT_val}(b)), E$^{3}$Relax-H$^{2}$ and EquiformerV2 produce more concentrated energy-difference distributions around zero compared with the other baselines. We note that the overall energy deviations on the vdW heterostructures dataset are slightly larger than those observed on C2DB. This can be attributed to the relatively small sample size and the broader chemical diversity of the vdW heterostructures, which introduces greater variability in the structural configurations.

Overall, these results confirm that the structural improvements achieved by E$^{3}$Relax-H$^{2}$ translate directly into energetically favorable configurations.

\begin{figure}[h]
  \centering
  \includegraphics[width=0.8\columnwidth]{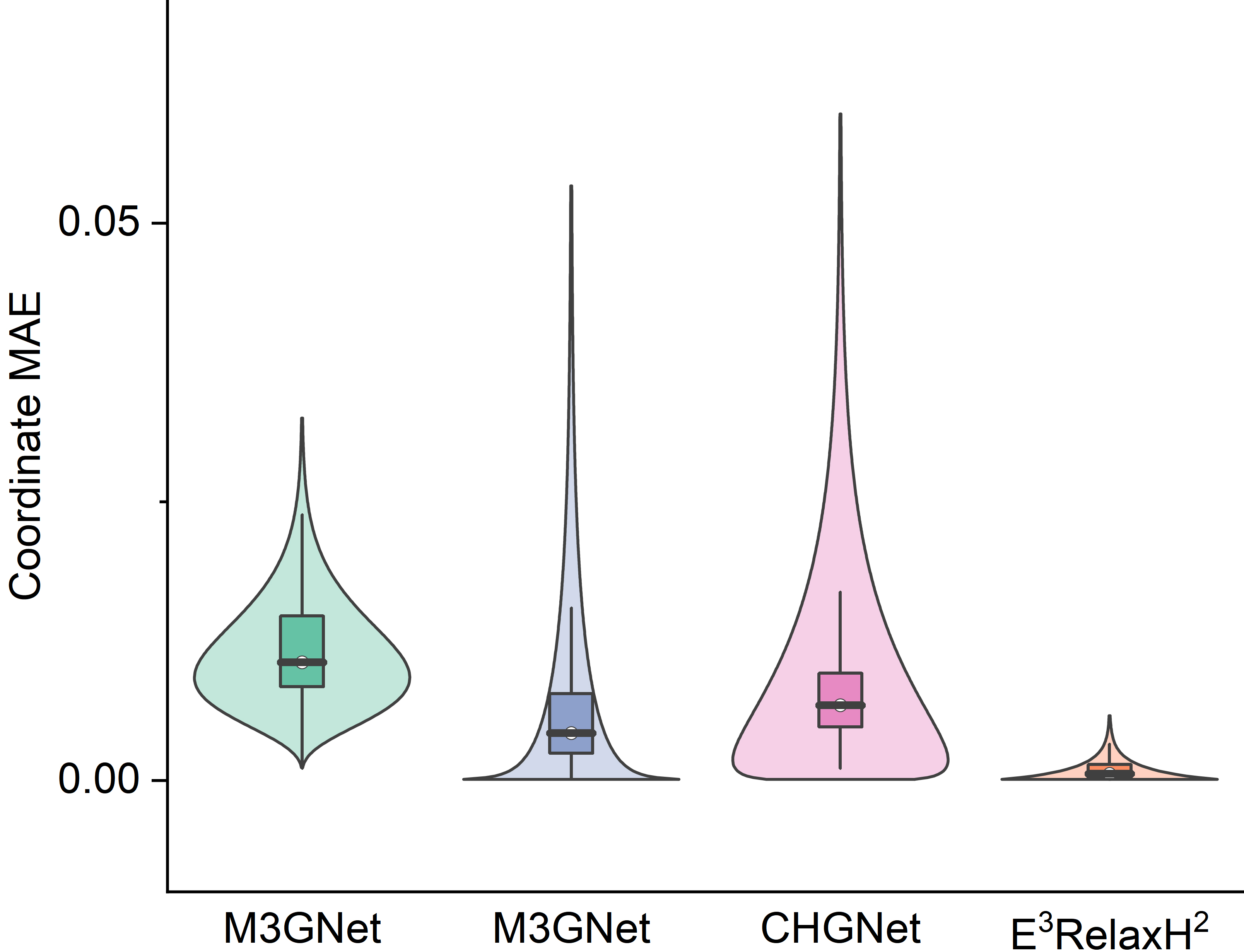}
  \caption{Comparison of coordinate MAE ($\rm \AA$) between E$^3$Relax and two ML-potential models using violin plots.}
  \label{fgr:MLIP}
\end{figure}

\subsection{Comparison with Iterative ML Models}
We further benchmark E$^{3}$Relax-H$^{2}$ against iterative ML-based potentials on a $\mathrm{MoS_2}$ point-defect dataset with energy, force, and stress labels. The dataset, compiled by Huang et al.~\cite{huang2023unveiling}, contains 5933 unique low-defect configurations in an $8\times8$ supercell, which we split into 4746 training, 593 validation, and 594 test samples. Using the original implementations, we train two representative MLIP, M3GNet~\cite{chen2022universal} and CHGNet~\cite{deng2023chgnet}.

As shown in Figure~\ref{fgr:MLIP}, E$^{3}$Relax-H$^{2}$ outperforms both M3GNet and CHGNet. We also observe that the iterative methods exhibit larger standard deviations in their coordinate errors, likely due to error accumulation over iterative optimization steps.

Iterative ML models have an important advantage in that they learn transferable quantities such as energies, forces, and stresses, which can support better cross-system generalization. However, this comes at the cost of more expensive data collection and slower inference. Moreover, stronger cross-system transferability does not necessarily translate into better performance on a specific target system, as demonstrated by our results on this dataset. On an NVIDIA L40S GPU, the average inference time per structure is \(3.72\pm5.13\) s for CHGNet, \(6.64\pm8.01\) s for M3GNet, and \(0.028\pm0.001\) s for E$^{3}$Relax-H$^{2}$.

\subsection{Ablation Study}
The effectiveness of E$^{3}$Relax-H$^{2}$ is attributed to three key strategies. First, the dual-node representation together with the lattice--atom message passing (LA-MP) module explicitly models the coupled evolution of atomic displacements and lattice deformations. Second, the proposed high-degree, high-order message passing (H$^2$-MP) module captures rich angular information and implicit many-body correlations while remaining computationally lightweight. Third, the differentiable periodicity-aware Cartesian displacement (PACD) loss resolves the ambiguity of Cartesian supervision under periodic boundary conditions. To assess the contribution of each component, we compare E$^{3}$Relax-H$^{2}$ with the following ablated variants:
\begin{itemize}
    \item \textbf{w/o lattice nodes}: We remove lattice nodes and predict relaxed lattice vectors from atom-node invariant features following the lattice-prediction procedure in DeepRelax~\cite{yang2024scalable}.
    \item \textbf{w/o H$^2$-MP}: We replace the H$^2$-MP module with the message-passing mechanism used in E$^{3}$Relax.
    \item \textbf{w/o PACD Loss}: We replace the PACD loss with the conventional displacement loss adopted in DeepRelax and E$^{3}$Relax.
\end{itemize}

Table~\ref{tbl:ablation_3D} reports results on MP and X-Mn-O, showing that each component contributes to the final performance. Notably, removing lattice nodes and predicting lattice vectors from atom features (as in DeepRelax~\cite{yang2024scalable}) breaks full SE(3)-equivariance because lattice updates are no longer represented and updated equivariantly within the graph. Consistent with this, the ablation results indicate that jointly preserving equivariance for both atomic coordinates and lattice vectors is important for improved accuracy.

\begin{table*}[!ht]
\centering
\caption{Ablation study on the two 3D materials datasets (MP and X-Mn-O). }
\begin{tabular}{lcccccc}
\toprule
\multirow{2}{*}{Model} &
\multicolumn{3}{c}{MP} &
\multicolumn{3}{c}{X-Mn-O} \\
\cmidrule(lr){2-4}\cmidrule(lr){5-7}
& MAE Coords.$\downarrow$ & Shape Dev.$\downarrow$ & MAE Vol.$\downarrow$
& MAE Coords.$\downarrow$ & Shape Dev.$\downarrow$ & MAE Vol.$\downarrow$ \\ \midrule
Dummy                         & 0.095 & 11.502 & 27.006 & 0.314 & 13.913 & 32.839 \\
w/o lattice nodes   & 0.058 &  4.546 &  8.312 & 0.106 &  3.741 &  3.467 \\
w/o H$^2$-MP           & 0.056 &  4.090 &  7.389 & 0.097 &  3.404 &  3.337 \\
w/o PACD loss & 0.055 &  3.937 &  7.368 & 0.096 &  3.353 &  3.323 \\
E$^{3}$Relax-H$^{2}$ 
& \textbf{0.054} & \textbf{3.904} & \textbf{7.307} 
& \textbf{0.092} & \textbf{3.265} & \textbf{3.191} \\
\bottomrule
\end{tabular}
\label{tbl:ablation_3D}
\end{table*}

\subsection{The Necessity of Equivariance}
Equivariance is not strictly required for crystal structure optimization. Nevertheless, since atomic coordinates, lattice vectors, and their relaxations transform in a predictable manner under rigid motions, enforcing $\mathrm{SE}(3)$-equivariance introduces a physically grounded inductive bias that improves data efficiency, generalization, and geometric consistency.

To examine its impact on data efficiency, we compare two variants of E$^{3}$Relax-H$^{2}$ on the relatively small vdW heterostructure dataset. The first variant does not enforce translation invariance in the lattice-to-atom broadcast stage, and uses $\hat{\mathbf{r}}_{ci}
={(\vec{\bm{l}}_c - \vec{\mathbf{r}}_i})/
{\|\vec{\bm{l}}_c - \vec{\mathbf{r}}_i\|}$. The second variant enforces translation invariance by replacing $\vec{\mathbf{r}}_i$ with the centered coordinate $\tilde{\vec{\mathbf{r}}}_i$, i.e., $\hat{\mathbf{r}}_{ci}
= {(\vec{\bm{l}}_c - \tilde{\vec{\mathbf{r}}}_i)} /
{\|\vec{\bm{l}}_c - \tilde{\vec{\mathbf{r}}}_i\|}$.

The experimental results show that the translation-invariant variant reduces the coordinate MAE from 0.120 to 0.114, suggesting that incorporating the correct symmetry prior is beneficial, especially in low-data regimes.

\subsection{Sensitivity Analysis of Key Hyperparameters}
We conduct a sensitivity analysis of key hyperparameters on the X-Mn-O dataset, including the number of layers ($T$), hidden dimension size ($F$), the maximum degree $\ell_{\max}$ used in the H$^2$-MP module, and the layer weighting scheme ($\alpha_t$). 

\begin{figure}[ht]
  \centering
  \includegraphics[width=1.0\columnwidth]{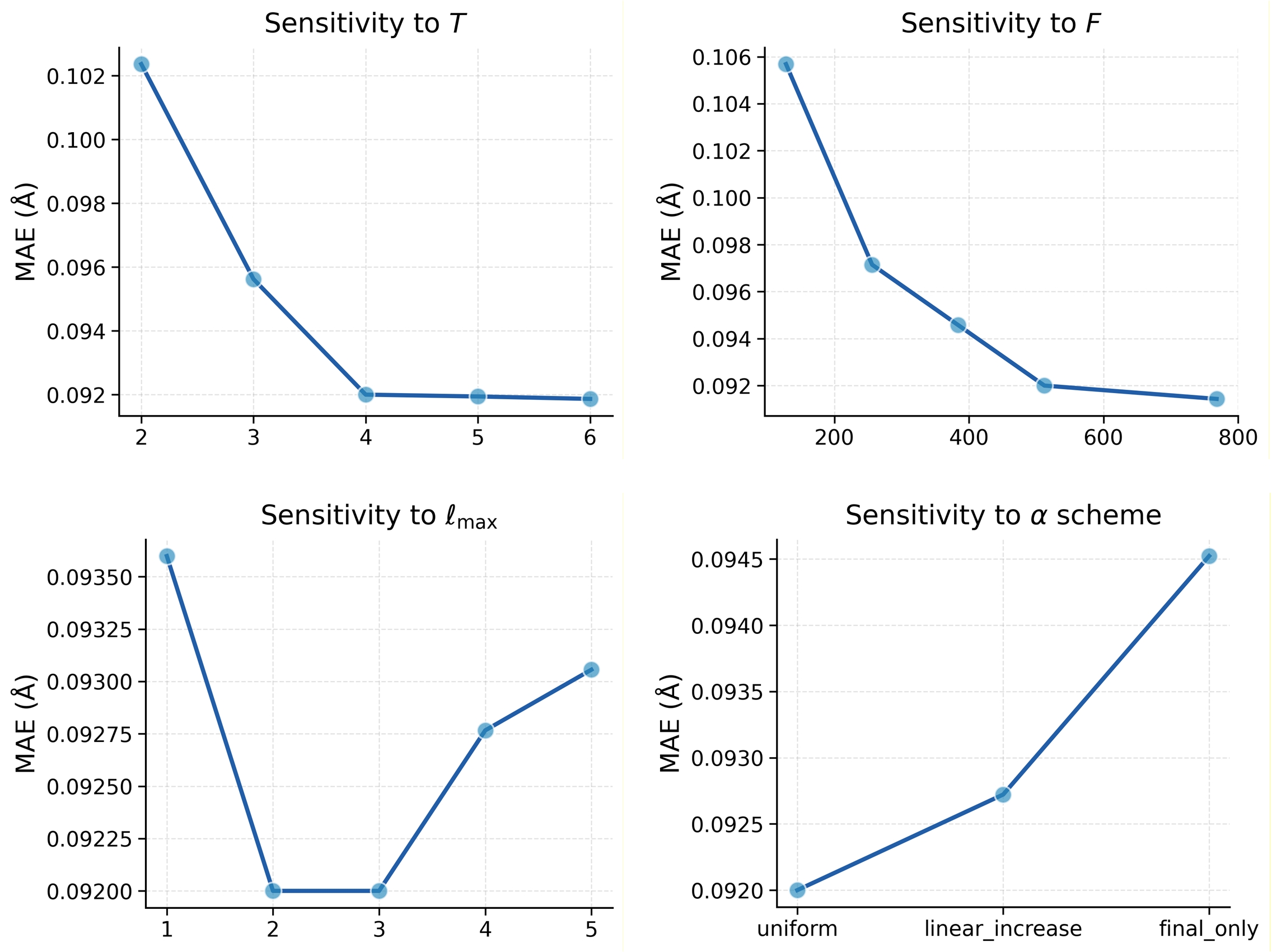}
  \caption{Sensitivity analysis of key hyperparameters on the X-Mn-O dataset, measured by coordinate MAE ($\rm \AA$). }
  \label{fgr:sensitivity}
\end{figure}

The results are presented in Figure~\ref{fgr:sensitivity}. Overall, these results indicate that E$^{3}$Relax-H$^{2}$ is robust to variations in key architectural hyperparameters and achieves stable performance across a wide range of configurations.

\subsection{Complexity Analysis}

Consider a crystal containing \(N\) atoms, each connected to \(K\) neighbors, and a hidden feature dimension \(F\). Lightweight equivariant GNNs such as PAINN and EGNN require \(\mathcal{O}(N K F^2)\) operations per layer, which arises from edge-wise message computation followed by feature mixing.

In E$^{3}$Relax-H$^{2}$, additional computational cost is introduced by two components, i.e., the high-degree high-order message passing module (H$^2$-MP) and the lattice-atom module.

\textbf{H$^2$-MP module.}  
For each node, the reference direction 
\(\hat{\mathbf{r}}_{\mathrm{ref},i}\) is computed by averaging neighbor directions, costing \(\mathcal{O}(N K)\).  
Spherical harmonics up to degree \(\ell_{\max}\) are then evaluated for each edge direction and reference direction, requiring \(\mathcal{O}(N K \ell_{\max}^2)\) operations.  
The bilinear contraction that produces invariants \(z^\ell_{ji}\) adds \(\mathcal{O}(N K \ell_{\max})\) cost.  
Since \(\ell_{\max}\) is a small constant (set to 3 in our experiments), these steps contribute only \(\mathcal{O}(N K)\) additional complexity.  
The dominant cost of the H$^2$-MP module remains the edge-wise MLPs and feature projections used to construct messages, which scale as \(\mathcal{O}(N K F^2)\).

\textbf{Lattice-atom module.}  
The lattice-atom module introduces interactions between the three lattice nodes and all atoms, costing \(\mathcal{O}(3 N F^2)\) for lattice-to-atom broadcast and another \(\mathcal{O}(3 N F^2)\) for atom-to-lattice aggregation. Lattice-lattice mixing adds only \(\mathcal{O}(F^2)\). 

Combining all components yields a total per-layer complexity of $\mathcal{O}\bigl(N (K+6) F^2\bigr)$. Overall, the added lattice-atom interactions introduce an extra $6NF^2$ term, which is about $12\%$ of the dominant $NKF^2$ term when $K \approx 50$. In practice, wall-clock overhead can be larger due to constant factors and implementation costs.

\section{Conclusion}
In this work, we presented E$^{3}$Relax-H$^{2}$, an end-to-end equivariant graph neural network for iteration-free crystal structure optimization. The proposed method addresses key limitations of existing approaches, including the lack of explicit lattice modeling, the difficulty of efficiently capturing high-degree angular information and high-order geometric correlations. Experiments on six benchmark datasets spanning both 2D and 3D materials show that E$^{3}$Relax-H$^{2}$ achieves state-of-the-art or highly competitive structural accuracy while maintaining high computational efficiency. DFT validation further confirms the physical validity of the predicted structures. Ablation and complexity analyses show that these gains come from the combined benefits of the novel designs, with only modest additional computational cost. These results establish E$^{3}$Relax-H$^{2}$ as an effective framework for one-shot crystal structure optimization.

\ifCLASSOPTIONcaptionsoff
  \newpage
\fi



\bibliographystyle{IEEEtran}
\bibliography{reference}

@article{gasteiger2021gemnet,
  title={Gemnet: Universal directional graph neural networks for molecules},
  author={Gasteiger, Johannes and Becker, Florian and G{\"u}nnemann, Stephan},
  journal={Advances in Neural Information Processing Systems},
  volume={34},
  pages={6790--6802},
  year={2021}
}

@article{haghighatlari2022newtonnet,
  title={Newtonnet: A newtonian message passing network for deep learning of interatomic potentials and forces},
  author={Haghighatlari, Mojtaba and Li, Jie and Guan, Xingyi and Zhang, Oufan and Das, Akshaya and Stein, Christopher J and Heidar-Zadeh, Farnaz and Liu, Meili and Head-Gordon, Martin and Bertels, Luke and others},
  journal={Digital Discovery},
  volume={1},
  number={3},
  pages={333--343},
  year={2022},
  publisher={Royal Society of Chemistry}
}

@article{xie2018crystal,
  title={Crystal graph convolutional neural networks for an accurate and interpretable prediction of material properties},
  author={Xie, Tian and Grossman, Jeffrey C},
  journal={Physical review letters},
  volume={120},
  number={14},
  pages={145301},
  year={2018},
  publisher={APS}
}

@article{schutt2017schnet,
  title={Schnet: A continuous-filter convolutional neural network for modeling quantum interactions},
  author={Sch{\"u}tt, Kristof and Kindermans, Pieter-Jan and Sauceda Felix, Huziel Enoc and Chmiela, Stefan and Tkatchenko, Alexandre and M{\"u}ller, Klaus-Robert},
  journal={Advances in neural information processing systems},
  volume={30},
  year={2017}
}

@inproceedings{gasteiger_dimenet_2020,
  title = {Directional Message Passing for Molecular Graphs},
  author = {Gasteiger, Johannes and Gro{\ss}, Janek and G{\"u}nnemann, Stephan},
  booktitle={International Conference on Learning Representations (ICLR)},
  year = {2020}
}

@article{choudhary2021atomistic,
  title={Atomistic line graph neural network for improved materials property predictions},
  author={Choudhary, Kamal and DeCost, Brian},
  journal={npj Computational Materials},
  volume={7},
  number={1},
  pages={185},
  year={2021},
  publisher={Nature Publishing Group UK London}
}

@inproceedings{schutt2021equivariant,
  title={Equivariant message passing for the prediction of tensorial properties and molecular spectra},
  author={Sch{\"u}tt, Kristof and Unke, Oliver and Gastegger, Michael},
  booktitle={International Conference on Machine Learning},
  pages={9377--9388},
  year={2021},
  organization={PMLR}
}

@inproceedings{liu2022spherical,
  title={Spherical message passing for 3d molecular graphs},
  author={Liu, Yi and Wang, Limei and Liu, Meng and Lin, Yuchao and Zhang, Xuan and Oztekin, Bora and Ji, Shuiwang},
  booktitle={International Conference on Learning Representations (ICLR)},
  year={2022}
}

@inproceedings{zitnick_scn_2022,
  title = {{Spherical Channels for Modeling Atomic Interactions}},
  author = {Zitnick, C. Lawrence and Das, Abhishek and Kolluru, Adeesh and Lan, Janice and Shuaibi, Muhammed and Sriram, Anuroop and Ulissi, Zachary and Wood, Brandon},
  booktitle = {Advances in Neural Information Processing Systems (NeurIPS)},
  year = {2022},
}

@article{chanussot2021open,
  title={Open catalyst 2020 (OC20) dataset and community challenges},
  author={Chanussot, Lowik and Das, Abhishek and Goyal, Siddharth and Lavril, Thibaut and Shuaibi, Muhammed and Riviere, Morgane and Tran, Kevin and Heras-Domingo, Javier and Ho, Caleb and Hu, Weihua and others},
  journal={Acs Catalysis},
  volume={11},
  number={10},
  pages={6059--6072},
  year={2021},
  publisher={ACS Publications}
}

@article{chen2022universal,
  title={A universal graph deep learning interatomic potential for the periodic table},
  author={Chen, Chi and Ong, Shyue Ping},
  journal={Nature Computational Science},
  volume={2},
  number={11},
  pages={718--728},
  year={2022},
  publisher={Nature Publishing Group US New York}
}

@article{
fu2023forces,
title={Forces are not Enough: Benchmark and Critical Evaluation for Machine Learning Force Fields with Molecular Simulations},
author={Xiang Fu and Zhenghao Wu and Wujie Wang and Tian Xie and Sinan Keten and Rafael Gomez-Bombarelli and Tommi S. Jaakkola},
journal={Transactions on Machine Learning Research},
issn={2835-8856},
year={2023},
url={https://openreview.net/forum?id=A8pqQipwkt},
}

@article{batzner20223,
  title={E (3)-equivariant graph neural networks for data-efficient and accurate interatomic potentials},
  author={Batzner, Simon and Musaelian, Albert and Sun, Lixin and Geiger, Mario and Mailoa, Jonathan P and Kornbluth, Mordechai and Molinari, Nicola and Smidt, Tess E and Kozinsky, Boris},
  journal={Nature communications},
  volume={13},
  number={1},
  pages={2453},
  year={2022},
  publisher={Nature Publishing Group UK London}
}

@inproceedings{gilmer2017neural,
  title={Neural message passing for quantum chemistry},
  author={Gilmer, Justin and Schoenholz, Samuel S and Riley, Patrick F and Vinyals, Oriol and Dahl, George E},
  booktitle={International conference on machine learning},
  pages={1263--1272},
  year={2017},
  organization={PMLR}
}

@article{fuchs2020se,
  title={Se (3)-transformers: 3d roto-translation equivariant attention networks},
  author={Fuchs, Fabian and Worrall, Daniel and Fischer, Volker and Welling, Max},
  journal={Advances in neural information processing systems},
  volume={33},
  pages={1970--1981},
  year={2020}
}

@inproceedings{brandstetter2021geometric,
  title={Geometric and Physical Quantities improve E (3) Equivariant Message Passing},
  author={Brandstetter, Johannes and Hesselink, Rob and van der Pol, Elise and Bekkers, Erik J and Welling, Max},
  booktitle={International Conference on Learning Representations},
  year={2021}
}

@inproceedings{satorras2021n,
  title={E (n) equivariant graph neural networks},
  author={Satorras, V{\i}ctor Garcia and Hoogeboom, Emiel and Welling, Max},
  booktitle={International conference on machine learning},
  pages={9323--9332},
  year={2021},
  organization={PMLR}
}

@article{shuaibi2021rotation,
  title={Rotation invariant graph neural networks using spin convolutions},
  author={Shuaibi, Muhammed and Kolluru, Adeesh and Das, Abhishek and Grover, Aditya and Sriram, Anuroop and Ulissi, Zachary and Zitnick, C Lawrence},
  journal={arXiv preprint arXiv:2106.09575},
  year={2021}
}

@article{batzner2023advancing,
  title={Advancing molecular simulation with equivariant interatomic potentials},
  author={Batzner, Simon and Musaelian, Albert and Kozinsky, Boris},
  journal={Nature Reviews Physics},
  pages={1--2},
  year={2023},
  publisher={Nature Publishing Group UK London}
}

@article{musaelian2023learning,
  title={Learning local equivariant representations for large-scale atomistic dynamics},
  author={Musaelian, Albert and Batzner, Simon and Johansson, Anders and Sun, Lixin and Owen, Cameron J and Kornbluth, Mordechai and Kozinsky, Boris},
  journal={Nature Communications},
  volume={14},
  number={1},
  pages={579},
  year={2023},
  publisher={Nature Publishing Group UK London}
}

@article{deng2023chgnet,
  title={CHGNet as a pretrained universal neural network potential for charge-informed atomistic modelling},
  author={Deng, Bowen and Zhong, Peichen and Jun, KyuJung and Riebesell, Janosh and Han, Kevin and Bartel, Christopher J and Ceder, Gerbrand},
  journal={Nature Machine Intelligence},
  pages={1--11},
  year={2023},
  publisher={Nature Publishing Group UK London}
}

@article{ong2013python,
  title={Python Materials Genomics (pymatgen): A robust, open-source python library for materials analysis},
  author={Ong, Shyue Ping and Richards, William Davidson and Jain, Anubhav and Hautier, Geoffroy and Kocher, Michael and Cholia, Shreyas and Gunter, Dan and Chevrier, Vincent L and Persson, Kristin A and Ceder, Gerbrand},
  journal={Computational Materials Science},
  volume={68},
  pages={314--319},
  year={2013},
  publisher={Elsevier}
}

@article{lyngby2022data,
  title={Data-driven discovery of 2D materials by deep generative models},
  author={Lyngby, Peder and Thygesen, Kristian Sommer},
  journal={npj Computational Materials},
  volume={8},
  number={1},
  pages={232},
  year={2022},
  publisher={Nature Publishing Group UK London}
}

@article{gibson2022data,
  title={Data-augmentation for graph neural network learning of the relaxed energies of unrelaxed structures},
  author={Gibson, Jason and Hire, Ajinkya and Hennig, Richard G},
  journal={npj Computational Materials},
  volume={8},
  number={1},
  pages={211},
  year={2022},
  publisher={Nature Publishing Group UK London}
}

@article{kim2023structure,
  title={A structure translation model for crystal compounds},
  author={Kim, Sungwon and Noh, Juhwan and Jin, Taewon and Lee, Jaewan and Jung, Yousung},
  journal={npj Computational Materials},
  volume={9},
  number={1},
  pages={142},
  year={2023},
  publisher={Nature Publishing Group UK London}
}

@article{yoon2020differentiable,
  title={Differentiable optimization for the prediction of ground state structures (DOGSS)},
  author={Yoon, Junwoong and Ulissi, Zachary W},
  journal={Physical Review Letters},
  volume={125},
  number={17},
  pages={173001},
  year={2020},
  publisher={APS}
}

@article{jain2013commentary,
  title={Commentary: The Materials Project: A materials genome approach to accelerating materials innovation},
  author={Jain, Anubhav and Ong, Shyue Ping and Hautier, Geoffroy and Chen, Wei and Richards, William Davidson and Dacek, Stephen and Cholia, Shreyas and Gunter, Dan and Skinner, David and Ceder, Gerbrand and others},
  journal={APL materials},
  volume={1},
  number={1},
  year={2013},
  publisher={AIP Publishing}
}

@article{gjerding2021recent,
  title={Recent progress of the computational 2D materials database (C2DB)},
  author={Gjerding, Morten Niklas and Taghizadeh, Alireza and Rasmussen, Asbj{\o}rn and Ali, Sajid and Bertoldo, Fabian and Deilmann, Thorsten and Kn{\o}sgaard, Nikolaj R{\o}rb{\ae}k and Kruse, Mads and Larsen, Ask Hjorth and Manti, Simone and others},
  journal={2D Materials},
  volume={8},
  number={4},
  pages={044002},
  year={2021},
  publisher={IOP Publishing}
}

@article{yan2022periodic,
  title={Periodic graph transformers for crystal material property prediction},
  author={Yan, Keqiang and Liu, Yi and Lin, Yuchao and Ji, Shuiwang},
  journal={Advances in Neural Information Processing Systems},
  volume={35},
  pages={15066--15080},
  year={2022}
}

@article{yang2025efficient,
  title={Efficient equivariant model for machine learning interatomic potentials},
  author={Yang, Ziduo and Wang, Xian and Li, Yifan and Lv, Qiujie and Chen, Calvin Yu-Chian and Shen, Lei},
  journal={npj Computational Materials},
  volume={11},
  number={1},
  pages={49},
  year={2025},
  publisher={Nature Publishing Group UK London}
}

@article{yang2024scalable,
  title={Scalable crystal structure relaxation using an iteration-free deep generative model with uncertainty quantification},
  author={Yang, Ziduo and Zhao, Yi-Ming and Wang, Xian and Liu, Xiaoqing and Zhang, Xiuying and Li, Yifan and Lv, Qiujie and Chen, Calvin Yu-Chian and Shen, Lei},
  journal={Nature Communications},
  volume={15},
  number={1},
  pages={8148},
  year={2024},
  publisher={Nature Publishing Group UK London}
}

@Article{Omee2024,
author={Omee, Sadman Sadeed
and Wei, Lai
and Hu, Ming
and Hu, Jianjun},
title={Crystal structure prediction using neural network potential and age-fitness Pareto genetic algorithm},
journal={Journal of Materials Informatics},
year={2024},
volume={4},
number={1},
pages={2},
doi={10.20517/jmi.2023.33}
}

@article{mosquera2024machine,
  title={Machine-learning structural reconstructions for accelerated point defect calculations},
  author={Mosquera-Lois, Irea and Kavanagh, Se{\'a}n R and Ganose, Alex M and Walsh, Aron},
  journal={npj Computational Materials},
  volume={10},
  number={1},
  pages={121},
  year={2024},
  publisher={Nature Publishing Group UK London}
}

@article{jiang2024machine,
  title={Machine learning potential assisted exploration of complex defect potential energy surfaces},
  author={Jiang, Chao and Marianetti, Chris A and Khafizov, Marat and Hurley, David H},
  journal={npj Computational Materials},
  volume={10},
  number={1},
  pages={21},
  year={2024},
  publisher={Nature Publishing Group UK London}
}

@article{huang2023unveiling,
  title={Unveiling the complex structure-property correlation of defects in 2D materials based on high throughput datasets},
  author={Huang, Pengru and Lukin, Ruslan and Faleev, Maxim and Kazeev, Nikita and Al-Maeeni, Abdalaziz Rashid and Andreeva, Daria V and Ustyuzhanin, Andrey and Tormasov, Alexander and Castro Neto, AH and Novoselov, Kostya S},
  journal={npj 2D Materials and Applications},
  volume={7},
  number={1},
  pages={6},
  year={2023},
  publisher={Nature Publishing Group UK London}
}

@article{batatia2022mace,
  title={MACE: Higher order equivariant message passing neural networks for fast and accurate force fields},
  author={Batatia, Ilyes and Kovacs, David P and Simm, Gregor and Ortner, Christoph and Cs{\'a}nyi, G{\'a}bor},
  journal={Advances in Neural Information Processing Systems},
  volume={35},
  pages={11423--11436},
  year={2022}
}

@article{
xu2024equivariant,
title={Equivariant Graph Network Approximations of High-Degree Polynomials for Force Field Prediction},
author={Zhao Xu and Haiyang Yu and Montgomery Bohde and Shuiwang Ji},
journal={Transactions on Machine Learning Research},
issn={2835-8856},
year={2024}
}

@inproceedings{
wang2025elora,
title={{EL}o{RA}: Low-Rank Adaptation for Equivariant {GNN}s},
author={Chen Wang and Siyu Hu and Guangming Tan and Weile Jia},
booktitle={Forty-second International Conference on Machine Learning},
year={2025}
}

@article{park2024scalable,
  title={Scalable parallel algorithm for graph neural network interatomic potentials in molecular dynamics simulations},
  author={Park, Yutack and Kim, Jaesun and Hwang, Seungwoo and Han, Seungwu},
  journal={Journal of chemical theory and computation},
  volume={20},
  number={11},
  pages={4857--4868},
  year={2024},
  publisher={ACS Publications}
}

@inproceedings{
    equiformer_v2,
    title={{EquiformerV2: Improved Equivariant Transformer for Scaling to Higher-Degree Representations}}, 
    author={Yi-Lun Liao and Brandon Wood and Abhishek Das and Tess Smidt},
    booktitle={International Conference on Learning Representations (ICLR)},
    year={2024}
}

@article{yang2025modeling,
  title={Modeling crystal defects using defect informed neural networks},
  author={Yang, Ziduo and Liu, Xiaoqing and Zhang, Xiuying and Huang, Pengru and Novoselov, Kostya S and Shen, Lei},
  journal={npj Computational Materials},
  volume={11},
  number={1},
  pages={229},
  year={2025},
  publisher={Nature Publishing Group UK London}
}

@article{cen2024high,
  title={Are high-degree representations really unnecessary in equivariant graph neural networks?},
  author={Cen, Jiacheng and Li, Anyi and Lin, Ning and Ren, Yuxiang and Wang, Zihe and Huang, Wenbing},
  journal={Advances in Neural Information Processing Systems},
  volume={37},
  pages={26238--26266},
  year={2024}
}

@inproceedings{aykent2025gotennet,
  title={Gotennet: Rethinking efficient 3d equivariant graph neural networks},
  author={Aykent, Sarp and Xia, Tian},
  booktitle={The Thirteenth International Conference on Learning Representations},
  year={2025}
}

@article{li2025critical,
  author    = {Yifan Li and Xiuying Zhang and Mingkang Liu and Lei Shen},
  title     = {A Critical Review of Machine Learning Interatomic Potentials and Hamiltonian},
  journal   = {Journal of Materials Informatics},
  year      = {2025},
  volume    = {5},
  number    = {4},
  pages     = {43},
  issn      = {2694-1015},
  doi       = {10.20517/jmi.2025.17},
  url       = {https://dx.doi.org/10.20517/jmi.2025.17},
}

@inproceedings{passaro2023reducing,
  title={Reducing SO (3) convolutions to SO (2) for efficient equivariant GNNs},
  author={Passaro, Saro and Zitnick, C Lawrence},
  booktitle={International conference on machine learning},
  pages={27420--27438},
  year={2023},
  organization={PMLR}
}

@inproceedings{du2022se,
  title={SE (3) equivariant graph neural networks with complete local frames},
  author={Du, Weitao and Zhang, He and Du, Yuanqi and Meng, Qi and Chen, Wei and Zheng, Nanning and Shao, Bin and Liu, Tie-Yan},
  booktitle={International Conference on Machine Learning},
  pages={5583--5608},
  year={2022},
  organization={PMLR}
}

@inproceedings{lin2023efficient,
  title={Efficient approximations of complete interatomic potentials for crystal property prediction},
  author={Lin, Yuchao and Yan, Keqiang and Luo, Youzhi and Liu, Yi and Qian, Xiaoning and Ji, Shuiwang},
  booktitle={International conference on machine learning},
  pages={21260--21287},
  year={2023},
  organization={PMLR}
}

@article{frank2024euclidean,
  title={A Euclidean transformer for fast and stable machine learned force fields},
  author={Frank, J Thorben and Unke, Oliver T and M{\"u}ller, Klaus-Robert and Chmiela, Stefan},
  journal={Nature Communications},
  volume={15},
  number={1},
  pages={6539},
  year={2024},
  publisher={Nature Publishing Group UK London}
}

@article{wen2024equivariant,
  title={An equivariant graph neural network for the elasticity tensors of all seven crystal systems},
  author={Wen, Mingjian and Horton, Matthew K and Munro, Jason M and Huck, Patrick and Persson, Kristin A},
  journal={Digital Discovery},
  volume={3},
  number={5},
  pages={869--882},
  year={2024},
  publisher={Royal Society of Chemistry}
}

@article{thomas2018tensor,
  title={Tensor field networks: Rotation-and translation-equivariant neural networks for 3d point clouds},
  author={Thomas, Nathaniel and Smidt, Tess and Kearnes, Steven and Yang, Lusann and Li, Li and Kohlhoff, Kai and Riley, Patrick},
  journal={arXiv preprint arXiv:1802.08219},
  year={2018}
}

@article{yang2024interaction,
  title={Interaction-based inductive bias in graph neural networks: enhancing protein-ligand binding affinity predictions from 3d structures},
  author={Yang, Ziduo and Zhong, Weihe and Lv, Qiujie and Dong, Tiejun and Chen, Guanxing and Chen, Calvin Yu-Chian},
  journal={IEEE Transactions on Pattern Analysis and Machine Intelligence},
  volume={46},
  number={12},
  pages={8191--8208},
  year={2024},
  publisher={IEEE}
}

@article{yin2025alphanet,
  title={AlphaNet: scaling up local-frame-based neural network interatomic potentials},
  author={Yin, Bangchen and Wang, Jiaao and Du, Weitao and Wang, Pengbo and Ying, Penghua and Jia, Haojun and Zhang, Zisheng and Du, Yuanqi and Gomes, Carla and Duan, Chenru and others},
  journal={npj Computational Materials},
  volume={11},
  number={1},
  pages={332},
  year={2025},
  publisher={Nature Publishing Group UK London}
}

@article{gong2023general,
  title={General framework for E (3)-equivariant neural network representation of density functional theory Hamiltonian},
  author={Gong, Xiaoxun and Li, He and Zou, Nianlong and Xu, Runzhang and Duan, Wenhui and Xu, Yong},
  journal={Nature Communications},
  volume={14},
  number={1},
  pages={2848},
  year={2023},
  publisher={Nature Publishing Group UK London}
}

@article{wen2025cartesian,
  title={Cartesian atomic moment machine learning interatomic potentials},
  author={Wen, Mingjian and Huang, Wei-Fan and Dai, Jin and Adhikari, Santosh},
  journal={npj Computational Materials},
  volume={11},
  number={1},
  pages={128},
  year={2025},
  publisher={Nature Publishing Group UK London}
}

@article{frank2022so3krates,
  title={So3krates: Equivariant attention for interactions on arbitrary length-scales in molecular systems},
  author={Frank, Thorben and Unke, Oliver and M{\"u}ller, Klaus-Robert},
  journal={Advances in Neural Information Processing Systems},
  volume={35},
  pages={29400--29413},
  year={2022}
}

@article{batatia2025design,
  title={The design space of E (3)-equivariant atom-centred interatomic potentials},
  author={Batatia, Ilyes and Batzner, Simon and Kov{\'a}cs, D{\'a}vid P{\'e}ter and Musaelian, Albert and Simm, Gregor NC and Drautz, Ralf and Ortner, Christoph and Kozinsky, Boris and Cs{\'a}nyi, G{\'a}bor},
  journal={Nature Machine Intelligence},
  volume={7},
  number={1},
  pages={56--67},
  year={2025},
  publisher={Nature Publishing Group UK London}
}

@ARTICLE{11098662,
  author={Liu, Wenjie and Zhu, Yifan and Zha, Ying and Wu, Qingshan and Jian, Lei and Liu, Zhihao},
  journal={IEEE Transactions on Pattern Analysis and Machine Intelligence}, 
  title={Rotation- and Permutation-Equivariant Quantum Graph Neural Network for 3D Graph Data}, 
  year={2025},
  volume={47},
  number={11},
  pages={10329-10343},
  doi={10.1109/TPAMI.2025.3593371}}

@article{han2025survey,
  title={A survey of geometric graph neural networks: Data structures, models and applications},
  author={Han, Jiaqi and Cen, Jiacheng and Wu, Liming and Li, Zongzhao and Kong, Xiangzhe and Jiao, Rui and Yu, Ziyang and Xu, Tingyang and Wu, Fandi and Wang, Zihe and others},
  journal={Frontiers of Computer Science},
  volume={19},
  number={11},
  pages={1911375},
  year={2025},
  publisher={Springer}
}

@article{liu2026traj2relax,
  title        = {Traj2Relax: A Trajectory-Supervised Method for Robust Structure Relaxation},
  author       = {Liu, Zhiyuan and Qian, Quan},
  journal      = {Journal of Chemical Information and Modeling},
  volume       = {66},
  number       = {4},
  pages        = {2081--2093},
  year         = {2026},
  publisher    = {American Chemical Society},
  doi          = {10.1021/acs.jcim.5c02703},
  url          = {https://doi.org/10.1021/acs.jcim.5c02703},
  issn         = {1549-9596}
}

@article{
lin2026learning,
title={Learning the energy relaxation manifold from unrelaxed structures with RelaxNet},
author={Emily Lin and Evelyn N. Wang and Sili Deng},
year={2026},
url={https://openreview.net/forum?id=2NZxmGjDZj}
}

@article{choudhary2020joint,
  title={The joint automated repository for various integrated simulations (JARVIS) for data-driven materials design},
  author={Choudhary, Kamal and Garrity, Kevin F and Reid, Andrew CE and DeCost, Brian and Biacchi, Adam J and Hight Walker, Angela R and Trautt, Zachary and Hattrick-Simpers, Jason and Kusne, A Gilad and Centrone, Andrea and others},
  journal={npj computational materials},
  volume={6},
  number={1},
  pages={173},
  year={2020},
  publisher={Nature Publishing Group UK London}
}

@inproceedings{yang2026equivariant,
  title={Equivariant Atomic and Lattice Modeling Using Geometric Deep Learning for Crystal Structure Optimization},
  author={Yang, Ziduo and Zhao, Yi-Ming and Wang, Xian and Zhuo, Wei and Liu, Xiaoqing and Shen, Lei},
  booktitle={Proceedings of the AAAI Conference on Artificial Intelligence},
  volume={40},
  number={33},
  pages={27747--27755},
  year={2026}
}



% \vfill

\enlargethispage{-5in}

\begin{IEEEbiography}[{\includegraphics[width=1in,height=1.25in,clip,keepaspectratio]{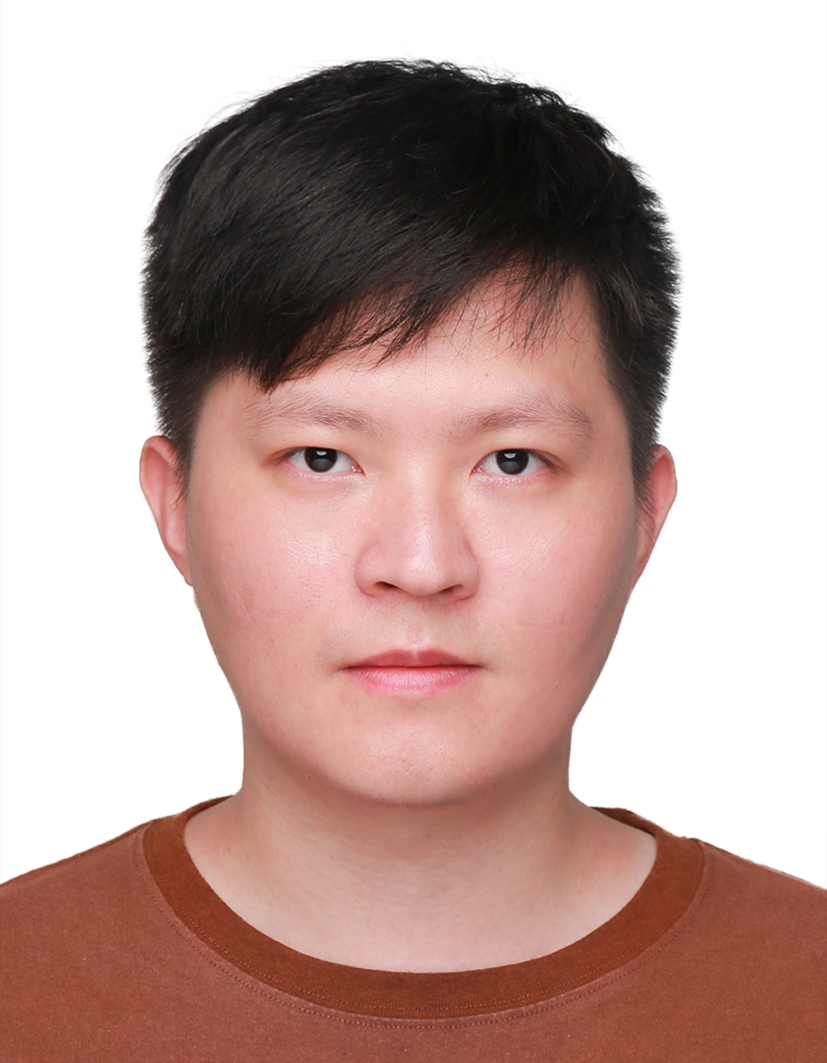}}]{Ziduo Yang}
is currently a lecturer with the College of Information Science and Technology, Jinan University, Guangzhou, China. He received the Ph.D. degree from Sun Yat-sen University, China. His research interests include AI for science, graph neural networks, and geometric deep learning.
\end{IEEEbiography}

\begin{IEEEbiography}[{\includegraphics[width=1in,height=1.25in,clip,keepaspectratio]{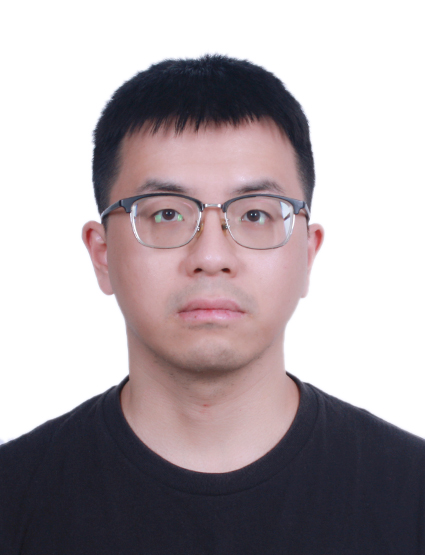}}]{Wei Zhuo}
is currently a research fellow with Nanyang Technological University, Singapore. He received the Ph.D. degree from Sun Yat-sen University, China. He was a visiting student in Computer Science at the University of Helsinki and the National University of Singapore. His research interests include machine learning, data mining, and graph representation learning.
\end{IEEEbiography}

\begin{IEEEbiography}[{\includegraphics[width=1in,height=1.25in,clip,keepaspectratio]{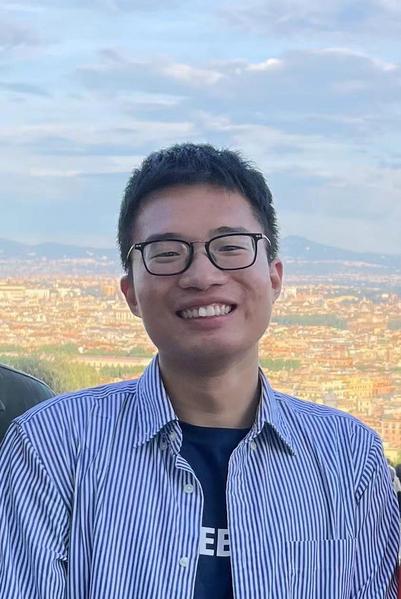}}]{Huiqiang Xie} received the B.S. degree from Northwestern Polytechnical University, the M.S. degree from Chongqing University, and the Ph.D. degree from Queen Mary University of London in 2023. From 2023 to 2024, he was a Postdoctoral Research Associate with the Hong Kong University of Science and Technology (Guangzhou). He is currently an Associate Professor with Jinan University. His research interests include deep learning and generative artificial intelligence.
\end{IEEEbiography}

\begin{IEEEbiography}[{\includegraphics[width=1in,height=1.25in,clip,keepaspectratio]{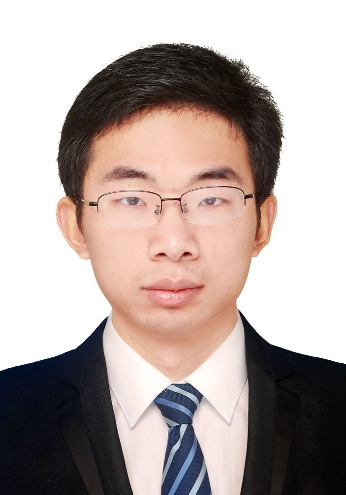}}]{Xiaoqing Liu} received the Ph.D. degree in Optical Engineering from Chongqing University, Chongqing, China, in 2023. He is currently a Postdoctoral Researcher with the Department of Mechanical Engineering, National University of Singapore, Singapore. His research focuses on the intersection of machine learning and computational materials science.
\end{IEEEbiography}

\begin{IEEEbiography}[{\includegraphics[width=1in,height=1.25in,clip,keepaspectratio]{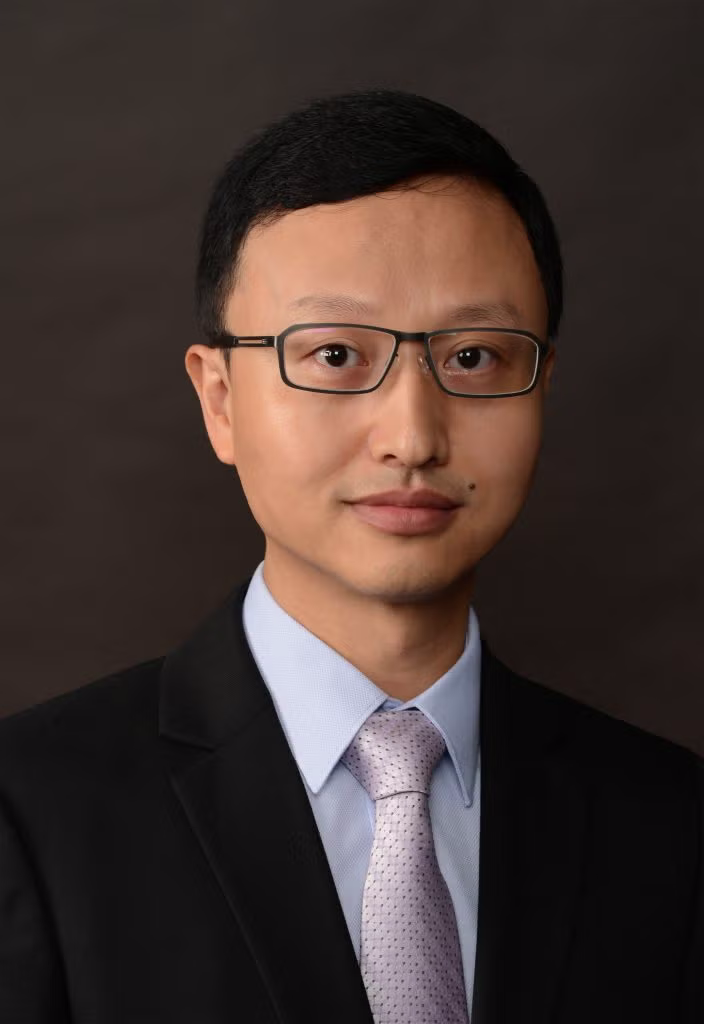}}]{Lei Shen} received the B.S. degree in Physics from Xiamen University and the Ph.D. degree in Computational Physics from the National University of Singapore. He is currently a Senior Lecturer in the Department of Mechanical Engineering at the National University of Singapore. His research focuses on the intersection of machine learning and computational materials science.
\end{IEEEbiography}

\end{document}